\documentclass[aps,pra,twocolumn,superscriptaddress]{revtex4}
\usepackage{bm}
\usepackage{graphicx}
\usepackage{amssymb,amsmath,amsbsy,amsgen,amsfonts}     
\usepackage{dcolumn}
\usepackage{amsthm}
\usepackage{mathrsfs}
\usepackage{latexsym}
\usepackage{array}
\usepackage{amstext}
\usepackage{epsfig}
\usepackage{epstopdf} % for TeXShop on Mark's Mac

\usepackage{color}
\usepackage{units}

\newcommand{\bra}[1]{\langle{#1}\vert}
\newcommand{\ket}[1]{\vert{#1}\rangle}
\newcommand{\ketbra}[2]{|#1\rangle \langle#2|}

\newcommand{\be}{\begin{equation}}
\newcommand{\ee}{\end{equation}}
\newcommand{\ba}{\begin{array}}
\newcommand{\ea}{\end{array}}
\newcommand{\bqa}{\begin{eqnarray}}
\newcommand{\eqa}{\end{eqnarray}}

\begin{document}

\title{Quantum plasmonic excitation in graphene and robust-to-loss propagation}

\author{George W. Hanson} 
\email{george@uwm.edu}
\address{Department of Electrical Engineering, University of Wisconsin-Milwaukee, 3200 N. Cramer St., Milwaukee, Wisconsin 53211, USA}

\author{S. A. Hassani Gangaraj} 
\address{Department of Electrical Engineering, University of Wisconsin-Milwaukee, 3200 N. Cramer St., Milwaukee, Wisconsin 53211, USA}

\author{Changhyoup Lee} 
\address{Centre for Quantum Technologies, National University of Singapore, 3 Science Drive 2,  117543 Singapore}

\author{Dimitris G. Angelakis} 
\address{Centre for Quantum Technologies, National University of Singapore, 3 Science Drive 2, 117543 Singapore}
\address{School of Electronic and Computer Engineering, Technical University of Crete, Chania, Greece 73100}

\author{Mark Tame} 
\email{markstame@gmail.com}
\address{University of KwaZulu-Natal, School of Chemistry and Physics, Durban 4001, South Africa}
\address{National Institute for Theoretical Physics, University of KwaZulu-Natal,Durban 4001,
South Africa}

\date{\today}

\begin{abstract}
We investigate the excitation of quantum plasmonic states of light in graphene using end-fire and prism coupling. In order to model the excitation process quantum mechanically we quantize the transverse-electric and transverse-magnetic surface plasmon polariton (SPP) modes in graphene. A selection of regimes are then studied that enable the excitation of SPPs by photons and we show that efficient coupling of photons to graphene SPPs is possible at the quantum level. Futhermore, we study the excitation of quantum states and their propagation under the effects of loss induced from the electronic degrees of freedom in the graphene. Here, we investigate whether it is possible to protect quantum information using quantum error correction techniques. We find that these techniques provide a robust-to-loss method for transferring quantum states of light in graphene over large distances.
\end{abstract}

\maketitle

%%%%%%%%%%%%%%%%%%%%%%%%%%%%%%%%%%%%%%%%%%%%%%
%%%%%%%%%%%%%%%%%%%%%%%%%%%%%%%%%%%%%%%%%%%%%%
%%%%%%%%%%%%%%%%%%%%%%%%%%%%%%%%%%%%%%%%%%%%%%

\section{Introduction}

Quantum plasmonics is attracting considerable interest at present from a wide range of researchers, most notably those in the quantum optics and plasmonics communities~\cite{Tame1}. This interest is in part due to the potential of plasmonics for applications in quantum information processing (QIP), which include ultracompact and versatile single-photon sources~\cite{Chang,Akimov,Kolesov,deLeon}, and single-photon switches~\cite{Chang2,Kolchin,Frank}. An important aspect in the quantum study of plasmonic systems is the excitation of quantum states of propagating surface plasmon polaritons (SPPs) using photons, which are easier to generate experimentally in a well controlled manner~\cite{OBrien}. Indeed, the ability to transfer a quantum state between two different systems is an important requirement for QIP in general~\cite{Di}. Here, the quantum state transfer must be efficient and not entail significant decoherence. Furthermore, the transferred state must be maintained over times and distances necessary to perform the required processing operations. Experimental work in quantum plasmonics using photons to excite SPPs has so far confirmed the preservation of entanglement when transferring quantum states between photons and SPPs~\cite{Alt,Fasel}, the preservation of superposition states~\cite{Fasel2}, as well as a wide range of other properties related to the quantum statistics of the excitation process~\cite{Ren,Ren2,Guo,Martino,Fujii}. Most recently experiments have demonstrated two-plasmon quantum interference, confirming the maintenance of the bosonic nature of the photons used to excite SPPs in a plasmonic Hong-Ou-Mandel setting~\cite{Fakonas, Martino2,Cai,Fujii2}. 
\begin{figure}[t]
\vskip0.5cm
\centerline{\includegraphics[width=8.3cm]{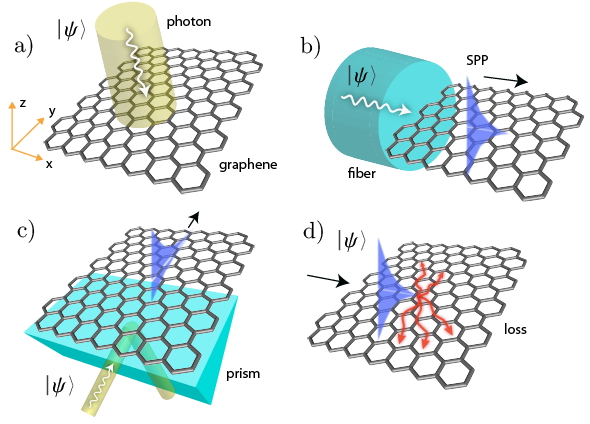}}
\caption{Excitation of quantum plasmonic states in graphene. {\bf (a)}:~Free-standing graphene and coordinate system. Here, SPPs cannot be excited by photons directly without a coupling method to provide energy-momentum matching conditions. {\bf (b)}:~End-fire method for coupling. {\bf (c)}:~Prism method for coupling. {\bf (d)}:~Using either end-fire or prism methods, the excited  plasmonic state propagates along the graphene surface. The hybrid nature of the SPP - consisting of a joint state of a photon and a collection of electrons - means that the electronic degrees of freedom induce loss effects in the photonic part. An error-correction code is introduced to deal with the loss and provide propagation over a large distance.}
\label{setup}
\end{figure} 

Most work on quantum plasmonic systems has so far focused on basic metallic material as the support media for the plasmonic excitations, using either silver or gold. However, there is a large range of other materials available to use~\cite{Tassin}, and graphene has recently emerged as a powerful alternative due to the possibility of chemical doping and electrical gating, which provides a highly tuneable media for supporting quantum plasmonic systems. Here, studies have focused on emitter coupling and decay into SPPs~\cite{Koppens}, active control over a quantum state biasing~\cite{Man}, nonlinear quantum optics~\cite{Gullans,Manz,Jablan}, quantum networks~\cite{Gullans2} and quantum sensing applications~\cite{Mus}. Despite some impressive work in this area, one key issue that has not been looked at in detail is the efficiency of the excitation of single SPPs in graphene using photons and more general quantum states of light. Moreover, given the many advantages that graphene offers compared to conventional plasmonic media in terms of tuneability, the excited SPPs still suffer from the effects of loss as they propagate.

In this work we investigate the two issues of the excitation efficiency and the effects of loss by studying the excitation of quantum plasmonic states of light in graphene using end-fire and prism coupling methods. We quantize the transverse-electric and transverse-magnetic SPP modes in graphene in order to build a fully quantum mechanical model for the excitation process. We then study various parameter regimes that enable the excitation of single SPPs by photons and find that efficient coupling of photons to graphene SPPs is possible. Furthermore, we study the subsequent propagation of the excited quantum states under the effects of loss induced from the electronic degrees of freedom in the graphene. In order to protect the quantum states from loss we use a quantum error-correction code and find that the code provides a robust-to-loss mechanism for propagating quantum states of light in graphene over large distances.

The work is divided into five sections. In Section II we introduce the model for SPP quantization in graphene. In Section III we then use this model to study the conversion of single photons to single graphene SPPs, and in Section IV we provide a model for describing the effects of loss during the subsequent propagation of the SPPs. In Section V we then investigate quantum state transfer and propagation in detail, introducing an error correction code for protecting against the effects of loss. We show the benefits of using the code compared to not using it. In Section VI we conclude with a summary of our results and an outlook on future studies.

%%%%%%%%%%%%%%%%%%%%%%%%%%%%%%%%%%%%%%%%%%%%%%
%%%%%%%%%%%%%%%%%%%%%%%%%%%%%%%%%%%%%%%%%%%%%%
%%%%%%%%%%%%%%%%%%%%%%%%%%%%%%%%%%%%%%%%%%%%%%

\section{Graphene SPP Quantization}

In our study of the transfer of quantum states between photons and graphene SPPs we consider the graphene as a free-standing sheet, as shown in Fig.~\ref{setup}~(a). Here, SPPs are excited by photons by using either the end of a fibre (end-fire method), as shown in Fig.~\ref{setup}~(b), or a prism, as shown in Fig.~\ref{setup}~(c). In order to model the coupling between photons and SPPs we must first quantize the SPPs in graphene. In this section we briefly summarize the quantization steps.

%%%%%%%%%%%%%%%%%%%%%%%%%%%%%%%%%%%%%%%%%%%%%%
%%%%%%%%%%%%%%%%%%%%%%%%%%%%%%%%%%%%%%%%%%%%%%

\subsubsection{Graphene Conductivity}

We start by considering a laterally-infinite graphene sheet lying in the $x-y$ plane, as shown in Fig.~\ref{setup}~(a). The graphene
is modeled as an infinitesimally-thin, local, two-sided surface characterized
by a surface conductivity $\sigma$~\cite{GSC2007},
\begin{align}
\sigma(  \omega)  =  &  \frac{ie^{2}k_{B}T}{\pi\hslash^{2}\left(
\omega+i\Gamma\right)  } \left(  \frac{\mu_{c}}{k_{B}T}+2\ln(
e^{-\frac{\mu_{c}}{k_{B}T}}+1)  \right) \label{GSC} \\
&  + \frac{i e^2(\omega+i\gamma)}{\pi \hbar^2} \int_{0}^{\infty}\frac{f_{d}(  -\varepsilon)
-f_{d}(  \varepsilon ) }{(\omega+i\gamma)^{2}-4( \varepsilon
/\hslash )^{2}}d\varepsilon \nonumber
\end{align}
where $\omega$ is the radian frequency, $\mu_{c}$ is the chemical potential (Fermi energy) and $\Gamma$ ($\gamma$) is a phenomenological intraband (interband) scattering rate. In addition, $T$ is the temperature, $e$ is the charge of an electron, $k_B$ is Boltzmann's constant and $f_{d}(  \varepsilon)  =(
e^{(  \varepsilon-\mu_{c})  /k_{B}T}+1)  ^{-1}$ is the
Fermi-Dirac distribution. The first term in Eq.~(\ref{GSC}) is due to intraband
contributions, while the second term is due to interband contributions. Absorption is associated with both intraband electron scattering and interband electron transitions. While Eq.~(\ref{GSC}) is valid for arbitrary $T$, when $k_{B}T\ll (\vert \mu_{c}\vert ,\hslash\omega)$, {\it i.e.}~the low temperature limit, it becomes
\cite{Koppens,Gus2}
\bqa
\sigma(  \omega)&=& \frac{ie^{2}\mu_{c}}{\pi\hslash^{2}(
\omega+i\Gamma)  }  \\
&& +\frac{e^{2}}{4\hslash}\bigg(  \Theta(  \hslash\omega-2\mu_{c})+\frac{i}{\pi}\ln \left\vert \frac{\hslash\omega-2\mu_{c}}{\hslash\omega+2\mu_{c}
} \right\vert  \bigg) \nonumber
\eqa
where $\Theta(  x)  $ is the Heaviside function. This form of the conductivity has no $T$ dependence and is used in our study to simplify calculations in the low temperature limit, whereas the full form given in Eq.~(\ref{GSC}) is used in the high temperature limit, {\it i.e.} room temperature ($T=300$K).

For the decay rates, typical intraband scattering times are $\tau=1/\Gamma=0.35$ ps at room
temperature, and as large as $\tau=3-5$ ps at low temperature \cite{Lee,Kim,Li,Tan}. For the interband scattering rate we use $1/\gamma=0.0658$ ps \cite{Daw}. These values are assumed throughout our work unless otherwise noted. The Drude form of the conductivity (the first term) has been verified in the far-infrared \cite{Kim,Li,Tan,Daw}, and in the near infrared and visible the interband behavior has
been verified \cite{Li}. Note that for the present application one could also
consider $N$ closely-spaced graphene monolayers, modeled as a single layer
with a larger effective conductivity, $\sigma_{\text{eff}}(  \omega)
=N\sigma(  \omega) $. 

%%%%%%%%%%%%%%%%%%%%%%%%%%%%%%%%%%%%%%%%%%%%%%
%%%%%%%%%%%%%%%%%%%%%%%%%%%%%%%%%%%%%%%%%%%%%%

\subsubsection{Classical Vector Potential}

We start by working in the Lorentz gauge ($\mathbf{E}=-\partial\mathbf{A/\partial t}$,
$\mathbf{B}=\nabla\times\mathbf{A}$) and consider a homogeneous material
having relative permittivity, $\varepsilon_{r}$, on either side of the graphene,
the wave equation for the vector potential is
\begin{equation}
\mathbf{\nabla}^{2}\mathbf{A}^{\pm}(  \mathbf{r},t)  -\frac
{1}{c^{2}}\frac{\partial^{2}}{\partial t^{2}}\mathbf{A}^{\pm}(
\mathbf{r},t)  = 0, \label{pe}%
\end{equation}
where $\pm$ corresponds to the region $z\gtrless0$ and $c=c_{0}/\sqrt{\varepsilon_{r}}$, with $c_{0}$ being
the speed of light in vacuum. The structure is invariant in the transverse
plane and therefore we can use the solution
\begin{equation}
\mathbf{A}^{\pm}(  \mathbf{r},t)  =\sum_{\mathbf{k}}\left[
\mathbf{A}_{\mathbf{k}}^{\pm}(  z)  e^{i(  \mathbf{k}%
\cdot\mathbf{r-\omega}_{\mathbf{k}}t)  }+\mathbf{A}_{\mathbf{k}}%
^{\pm\ast}(  z)  e^{-i(  \mathbf{k}\cdot\mathbf{r}%
-\omega_{\mathbf{k}}t)  }\right],  \label{a}%
\end{equation}
where the wave vector $\mathbf{k}=k_{x}\mathbf{x}+k_{y}\mathbf{y}$ is
parallel to the interface. Substituting Eq.~(\ref{a}) into Eq.~(\ref{pe}) we find
\begin{equation}
(  \frac{\partial^{2}}{\partial z^{2}}-q_{0}^{2})  \mathbf{A}%
_{\mathbf{k}}^{\pm}(  z)  =0,\ \ \
\end{equation}
where $q_{0}^{2}=k^{2}-\frac{1}{c^{2}}\omega_{\mathbf{k}}^{2}$, with
$k^{2}=\vert \mathbf{k}\vert ^{2}$. Thus, a solution for the vector potential is
\begin{equation}
\mathbf{A}^{\pm}(  \mathbf{r},t)  =\sum_{\mathbf{k}}\mathbf{A}
_{\mathbf{k}}^{\pm}  e^{\mp q_{0}z} e^{i(  \mathbf{k}\cdot
\mathbf{r-\omega}_{\mathbf{k}}t)  }+\text{c.c.} \label{vecpot}
\end{equation}
Assuming for the moment propagation in the $x$ direction only, {\it i.e.} invariance in the $y$-direction ($k_{y}=0$), for TE modes the non-zero field
components are ($E_{y}$, $H_{x}$, $H_{z}$) arising from $A_{y}$, and for TM
modes we have non-zero components ($E_{x}$, $E_{z}$, $H_{y}$) arising from $A_{x}$ and $A_{z}$.
Enforcing the boundary conditions~\cite{Landau}
\begin{align}
\mathbf{z}\times(  \mathbf{H}^{+}-\mathbf{H}^{-})   &
=\mathbf{J}=\sigma\mathbf{E}\label{BC}\\
\mathbf{z}\times(  \mathbf{E}^{+}-\mathbf{E}^{-})   &
=0\nonumber
\end{align}
at $z=0$ leads to the determination of the form of the ${\bf A}_{\mathbf k}^\pm$ vectors and to the dispersion relations for TE and TM graphene modes that link the wavenumber $k$ to the frequency $\omega_k$ via the conductivity, $\sigma(\omega_k)$, and the relative permittivity of the background material, $\varepsilon_r$. The explicit form of the ${\bf A}_{\mathbf k}^\pm$ vectors is given later in this section. First, we discuss the dispersion relations, which are given by
\begin{align}
k^{\text{TM}}=k_{x}^{\text{TM}}  &  =k_{0}\sqrt{\varepsilon_r \left(1-\left(  \frac{2}%
{\sigma\eta}\right)  ^{2}\right) },\label{kp}\\
k^{\text{TE}}=k_{x}^{\text{TE}}  &  =k_{0}\sqrt{\varepsilon_r \left(1-\left(  \frac
{\sigma\eta}{2}\right)  ^{2}\right)  },\label{kp2}
\end{align}
where $k_{0}=\omega_{k}/c_0$, $\eta=\sqrt{\mu_0\mu_r/\varepsilon_0\varepsilon_r}$ (with $\mu_r=1$) and
$\sigma=\sigma\left(  \omega_{k}\right)  $. 
\begin{figure}[t]
\centerline{\psfig{figure=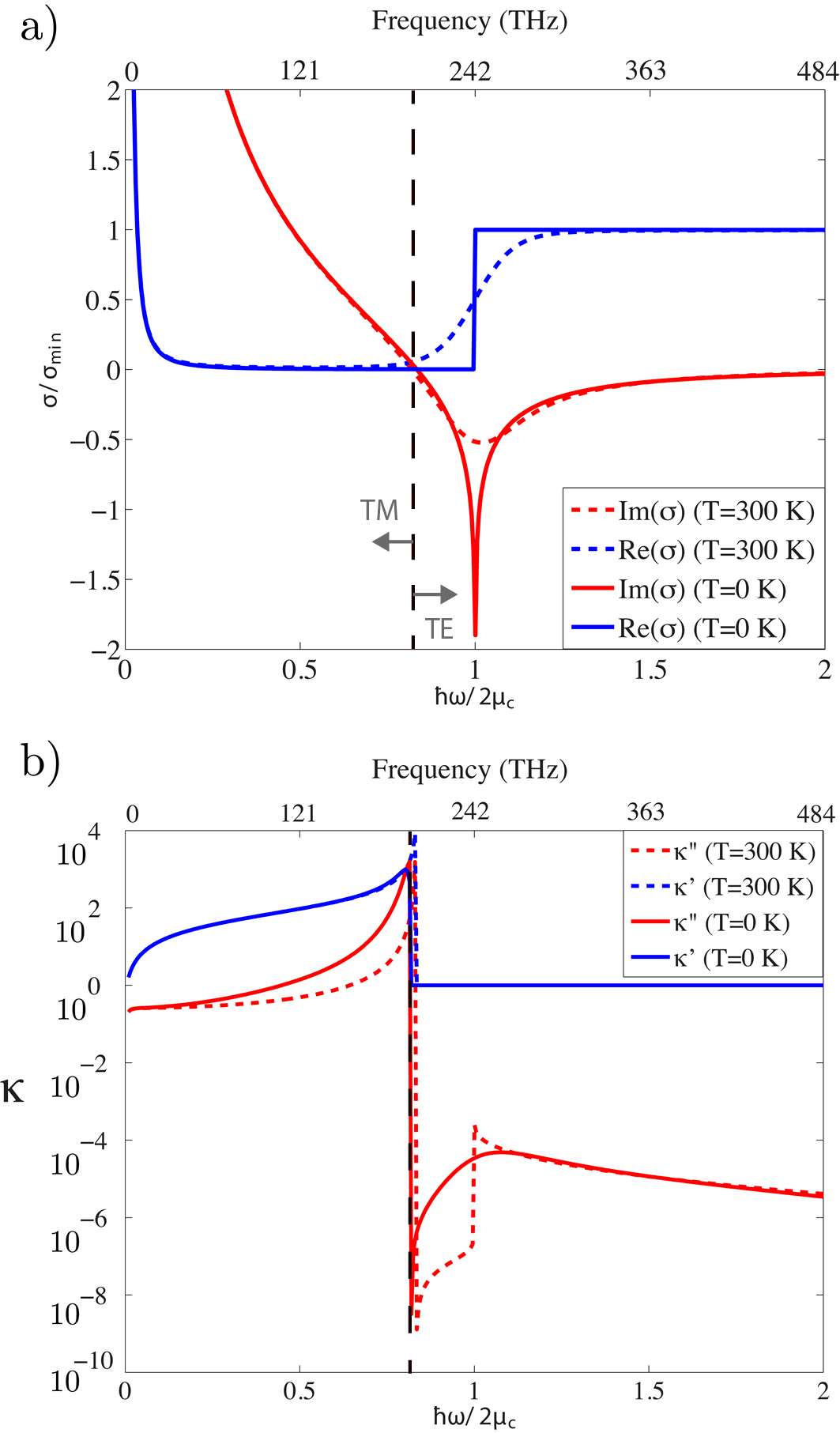,width=7.0cm}}
\caption{Graphene conductivity and surface plasma wavenumber. {\bf (a)}:~The conductivity $\sigma(\omega)$ for bare graphene. The conductivity has been rescaled by $\sigma_{\rm min}=\pi e^2/2h$. {\bf (b)}:~The rescaled surface plasma wavenumbers for TE and TM modes ($\varepsilon_r=1$), $\kappa=k_{x}/k_{0}=\kappa^{\prime}+i\kappa^{\prime\prime}$. In both, the chemical potential is $\mu_{c}=0.5$ eV, and we have set $\tau=5$~ps at $T=0$~K and $\tau=0.35$ ps at $T=300$~K. }
\label{conduc}
\end{figure} 

By decomposing the conductivity into real and imaginary parts, $\sigma
=\sigma^{\prime}+i\sigma^{\prime\prime}$, it can be shown that that for $\sigma^{\prime\prime}>0$
(inductive surface reactance) only a single TM surface plasma wave is supported by the graphene and can propagate; if $\sigma^{\prime\prime}<0$
(capacitive surface reactance) the TM SPP is on the improper Riemann sheet, exponentially increasing as $\lvert z \rvert \rightarrow \infty$ \cite{MZPRL2007,Han2008}. 

On the other hand, for $\sigma^{\prime\prime}<0$ (capacitive surface reactance) only a single TE surface plasma wave can propagate (if $\sigma^{\prime\prime}>0$
the TE SPP is on the improper Riemann sheet) \cite{MZPRL2007,Han2008}. Fig.~\ref{conduc}~(a) shows the
conductivity $\sigma$ and Fig.~\ref{conduc}~(b) shows the rescaled surface plasma wavenumber, $\kappa=k/k_0$ for both TE and TM modes, decomposed into real part $\kappa^\prime=\operatorname{Re}
(  k_{x}/k_{0})$ and imaginary part $\kappa^{\prime
\prime}=\operatorname{Im}(  k_{x}/k_{0})$ over a wide range of frequencies for $T=0$~K and $T=300$~K, with the chemical potential chosen as $\mu_{c}=0.5$ eV. At low frequencies $\hslash\omega\ll2\mu_{c}$ ($\omega=\omega_k$ is used for concise notation) the intraband
conductivity is dominant ($\sigma^{\prime\prime}\simeq\sigma_{\text{intra}
}^{\prime\prime}$), interband absorption is blocked, and a slow TM surface plasma wave
can propagate on the graphene surface ($\sigma^{\prime\prime}>0$). As the frequency increases the Drude term falls off, and in the vicinity of $\hslash\omega/2\mu_{c}=1$ interband absorption becomes important. As the frequency increases further $\sigma^{\prime\prime}\simeq\sigma
_{\text{inter}}^{\prime\prime}$, so that a loosely-bound TE surface plasma wave can propagate $(\sigma^{\prime\prime}<0)$. Therefore, in Fig. 2(b), the mode is TM to the left of the discontinuity (TE mode is on the improper Riemann sheet), and TE to the right of the discontinuity (TM mode is on the improper Riemann sheet). Note that the frequency range of the two modes can be adjusted by changing the chemical potential, which will shift the discontinuity to the left or right as required.

The vector potential from Eq.~(\ref{vecpot}) can be written explicitly as
\begin{equation}
\mathbf{A}^{\pm}({\bf r}, t)=\sum_{\mathbf{k}}C_{\mathbf{k}}{\bm \phi}_\mathbf{k}(
z )  e^{i(  \mathbf{k}\cdot\mathbf{r-\omega}_{\mathbf{k}%
}t)  }+\text{c.c.,} \label{vp}%
\end{equation}
where the $C_{\mathbf{k}}$ are mode amplitudes and the mode functions are
\begin{align}
\bm{\phi}_\mathbf{k}(  z) & =\bm{\phi}^{+}_\mathbf{k}(
z )  +\bm{\phi}^{-}_\mathbf{k}( z )
,\label{MF}\\
\bm{\phi}_{\mathbf{k}, \text{TM}}^{\pm} (  z )   &  =-\Theta
(  \pm z)  \left( \frac{2k_{x}i}
{2q_{0}-i\sigma\mu_0\omega_{\mathbf{k}}}\hat{\mathbf{x}}\mp\hat{\mathbf{z}}\right)  e^{\mp
q_{0}z} \label{MF2}\\
\bm{\phi}_{\mathbf{k}, \text{TE}}^{\pm} (  z )   &  =\Theta(
\pm z)  \hat{\mathbf{y}}e^{\mp q_{0}z}. \label{MF3}
\end{align}
Note that, from the relation $q_{0}^{2}=k^{2}-\frac{1}{c^{2}}\omega_{\mathbf{k}
}^{2}$, and Eqs.~(\ref{kp})~and~(\ref{kp2}), we have
\begin{equation}
q_{0}^{\text{TM}}=\pm i \frac{\omega_{\mathbf{k}}}{c}\frac{2}{\sigma
\eta},\ \ q_{0}^{\text{TE}}=\pm i \frac{\omega_{\mathbf{k}}}%
{c}\frac{\sigma\eta}{2},
\end{equation}
with the sign chosen so that the modal field decays exponentially in the vertical direction away from the graphene sheet. As $\sigma(\omega)$ is a complex valued function, in general $q_{0}=q_{0}' + i q_{0}''$, where $q_{0}'$ describes the strength of lateral field confinement (the decay length) to the graphene sheet while $q_{0}''$ corresponds to the wavelength of the propagating mode in the $z$-axis, corresponding to leakage into the far-field. For simplicity, in this work we focus on propagating modes in the $x-y$ plane only, so we restrict our interest to the regime where ${\rm Re}(q_{0}) \gg {\rm Im}(q_{0})$. Furthermore, our model will only include loss in $x-y$ plane, corresponding to the case where internal loss is much greater than leaky loss originating from $q_0''$.

%%%%%%%%%%%%%%%%%%%%%%%%%%%%%%%%%%%%%%%%%%%%%%
%%%%%%%%%%%%%%%%%%%%%%%%%%%%%%%%%%%%%%%%%%%%%%

\subsubsection{Quantization}

With the explicit form of the vector potential now given, we proceed to quantize the surface plasma wave. Note that here we are considering an ideal case with no damping effects due to the electronic degrees of freedom in the graphene sheet. Furthermore, and consistent with no damping, for quantization we assume a non-dispersive model since we consider quantum states that are associated with wavepackets that have very narrow bandwidths centered on $\omega$. Both assumptions simplify the quantization procedure. Damping will be reintroduced to the model later in Section IV. 

We start with the Hamiltonian for the field given by
\begin{equation}
H_{f}=\frac{\varepsilon_0}{2}\int_{V}\left(  \varepsilon_r \bigg\vert \frac{\partial
\mathbf{A}}{\partial t}\bigg\vert ^{2}+c_0^{2}\vert \nabla\times
\mathbf{A}\vert ^{2}\right)  dV.\nonumber
\end{equation}

For TE modes ($\sigma ^{\prime \prime }<0$) the energy stored in the
graphene can be obtained by temporarily assuming the graphene has a small
but non-zero thickness $d$ with effective permittivity 
\begin{equation}
\varepsilon =1+i\frac{\sigma }{\omega d}=1-\frac{\sigma ^{\prime \prime }}{%
\omega d}=1+\frac{\left\vert \sigma ^{\prime \prime }\right\vert }{\omega d}%
\simeq \frac{\left\vert \sigma ^{\prime \prime }\right\vert }{\omega d},
\label{eps1}
\end{equation}%
with associated energy%
\bqa
H_{e}^{\text{TE}}&=&\frac{1}{2}\int_{V}\varepsilon |\mathbf{E}|^{2}dV=\frac{1}{2}%
\int_{V}\frac{\left\vert \sigma ^{\prime \prime }\right\vert }{\omega d}%
|\mathbf{E}|^{2}dV \nonumber \\
&=&\frac{1}{2}\int_{V}\delta \left( z\right) \omega \left\vert
\sigma ^{\prime \prime }\right\vert |\mathbf{A}_{\Vert }|^{2}dV
\eqa
where $\mathbf{A}_{\Vert }$ is the component of potential parallel to the
graphene sheet. Note that at this stage we are not including loss, so that $\sigma=i\sigma''$.

For TM modes $\sigma ^{\prime \prime }>0$, leading to a negative
permittivity $\varepsilon <0$, and so a different method
must be used. We again assume that the graphene has a small but non-zero
thickness $d$. Inside the graphene the equation of motion $\mathbf{F}=m%
\mathbf{a}$ leads to%
\begin{equation}
\frac{\partial \mathbf{A}_{\Vert }}{\partial t}=\frac{m}{e}\frac{d\mathbf{v}%
}{dt},
\end{equation}%
where $\mathbf{v}$ is the velocity of electrons in the electron gas. Thus, $%
\mathbf{A}_{\Vert }\left( \mathbf{r},t\right) =\left( m/e\right) \mathbf{v}%
\left( \mathbf{r},t\right) $ with associated current density $\mathbf{J}%
=-en\left( \mathbf{r},t\right) \mathbf{v}\left( \mathbf{r},t\right) =-\left(
e^{2}/m\right) n\left( \mathbf{r},t\right) \mathbf{A}_{\Vert }\left( \mathbf{%
r},t\right) $, where $n$ is the number density. The energy stored in the
graphene electron gas kinetics is \cite{ER1971}
\bqa
H_{e}&=&\frac{1}{2}m\int n\left( \mathbf{r},t\right) \mathbf{v}^{2}\left( 
\mathbf{r},t\right) dV \nonumber \\
&=&\frac{1}{2}\frac{e^{2}}{m}\int_{V}n\left( \mathbf{r}%
,t\right) \mathbf{A}_{\Vert }^{2}\left( \mathbf{r},t\right) dV,
\eqa
Considering that for a lossless plasma $n_{e}=\omega _{p}^{2}\varepsilon
_{0}\left( m/e^{2}\right) =m\omega \sigma ^{\prime \prime }/e^{2}d$, we
obtain 
\begin{equation}
H_{e}^{\text{TM}}=\frac{1}{2}\int_{V}\delta \left( z\right) \omega \sigma
^{\prime \prime }\mathbf{A}_{\Vert }^{2}dV.
\end{equation}%
Therefore, the total Hamiltonian for either TE or TM modes is  
\bqa
H&=&\frac{\varepsilon _{0}}{2}\int dV\bigg( \varepsilon _{r}\left\vert \frac{%
\partial \mathbf{A}}{\partial t}\right\vert ^{2} \\
&& \hskip1.8cm +c_{0}^{2}\left\vert \nabla
\times \mathbf{A}\right\vert ^{2}+\delta \left( z\right) \frac{\omega
\left\vert \sigma ^{\prime \prime }\right\vert }{\varepsilon _{0}}\left\vert 
\mathbf{A}_{\Vert }\right\vert ^{2}\bigg). \nonumber
\eqa

To compute the Hamiltonian, we take a region of space of size $L\times L$ on the surface of the graphene sheet (in the $x-y$ plane) such that the wavenumbers are $k_{x,y}=2\pi
n_{x,y}/L$, $n_{x,y}=0,\pm1,\pm2,...$, {\it i.e.} periodic boundary conditions. The
volume integrals in the Hamiltonian are evaluated using $\int_{0}^{L}\int
_{0}^{L}e^{i(  \mathbf{k}-\mathbf{p})  \cdot\mathbf{r}}%
dxdy=L^{2}\delta_{\mathbf{k},\mathbf{p}}$ and substituting in Eq.~(\ref{vp}). By doing this we obtain
terms such as
\begin{align}
\vert \mathbf{A}(  \mathbf{r},t)\vert ^{2}   &
=2\sum_{\mathbf{k}}C_{\mathbf{k}}C_{\mathbf{k}}^{\ast}\bm{\phi}_\mathbf{k}(
\mathbf{r})  \cdot\bm{\phi}^{\ast}_\mathbf{k}( \mathbf{r}) \nonumber \\
&  +\sum_{\mathbf{k}}C_{\mathbf{k}}C_{-\mathbf{k}}\bm{\phi}_\mathbf{k}(\mathbf{r})  \cdot\bm{\phi}_{-\mathbf{k}}(\mathbf{r})  e^{-i2\omega_{\mathbf{k}}t}+\text{c.c.}\nonumber
\end{align}
It is straightforward to show that the cross terms, {\it e.g.}~$\bm{\phi
}_\mathbf{k}(\mathbf{r})\cdot\bm{\phi}_{-\mathbf{k}}(
\mathbf{r})  $, associated with electric, magnetic, and gas
kinetic energies will cancel, as occurs in free-space optics, and thus we only
retain terms such as $\bm{\phi}_\mathbf{k}(  \mathbf{r})
\cdot\bm{\phi}_\mathbf{k}^{\ast}(  \mathbf{r})  =\vert
\bm{\phi}_\mathbf{k}(  \mathbf{r})  \vert ^{2}$. By collecting all the non-zero terms, after substituting in Eq.~(\ref{vp}) and carrying out the integrals, we obtain
\begin{equation}
H=\varepsilon_{0}L^{2}\sum_{\mathbf{k}}\omega_{\mathbf{k}}^{2}N_{\mathbf{k}%
}(  C_{\mathbf{k}}^{\ast}C_{\mathbf{k}}+C_{\mathbf{k}}C_{\mathbf{k}%
}^{\ast}), \label{Hamspp}
\end{equation}
where $N_{\mathbf{k}}=N_{\mathbf{k}}^{\text{TE/TM}}$ is a normalization
parameter having units of length. To quantize the fields, we use the correspondence of the Hamiltonian in Eq.~(\ref{Hamspp}) with that of the harmonic
oscillator, mapping the coefficients as
\begin{equation}
C_{\mathbf{k}}\rightarrow\sqrt{\frac{\hslash}{2\varepsilon_0 L^{2}%
\omega_{\mathbf{k}}N_{\mathbf{k}}}}\hat{b}_{\mathbf{k}},\ \ C_{\mathbf{k}%
}^{\ast}\rightarrow\sqrt{\frac{\hslash}{2\varepsilon_0 L^{2}\omega_{\mathbf{k}%
}N_{\mathbf{k}}}}\hat{b}_{\mathbf{k}}^{\dagger},
\end{equation}
\noindent where the bosonic creation and annihilation operators satisfy the commutation relation $[
\hat{b}_{\mathbf{k}},\hat{b}_{\mathbf{k}^{\prime}}^{\dagger}]
=\delta_{\mathbf{kk}^{\prime}}$. We then have the quantized Hamiltonian, $\hat{H}=\sum_{\mathbf{k}}(
\hslash\omega_{\mathbf{k}}/2) (  \hat{b}_{\mathbf{k}}
\hat{b}_{\mathbf{k}}^{\dagger}+\hat{b}_{\mathbf{k}}^{\dagger}
\hat{b}_{\mathbf{k}})$, for the system and the corresponding quantized vector potential operator is 
\begin{equation}
\hat{\mathbf{A}}(  \mathbf{r},t)  =\sum_{\mathbf{k}}\sqrt
{\frac{\hslash}{2\varepsilon_0 L^{2}\omega_{\mathbf{k}}N_{\mathbf{k}}}
}\hat{b}_{\mathbf{k}}\bm{\phi}_\mathbf{k}( z)
e^{i(  \mathbf{k}\cdot\mathbf{r-\omega}_{\mathbf{k}}t)
}+\text{H.c.}%
\end{equation}
For our purposes it is convenient to take the continuum limit using $\sum_{\mathbf{k}}%
\rightarrow(  \frac{L}{2\pi})  ^{2}\int d\mathbf{k}$ and $ \hat{b}_{\mathbf{k}}\rightarrow(  \frac{2\pi}{L})  \hat
{b}(  \mathbf{k})  $, so that
\begin{equation}
\hat{\mathbf{A}}( \mathbf{r},t)  =\frac{1}{2\pi}\int
d\mathbf{k}\sqrt{\frac{\hslash}{2\varepsilon_0\omega N({\mathbf{k}})}%
}\bm{\phi}( z,\mathbf{k})  e^{i(  \mathbf{k}%
\cdot\mathbf{r-\omega} t)  }\hat{b}(
\mathbf{k})  +\text{H.c.}%
\end{equation}
For simplicity, we also assume excitations propagating in the $x$-direction only with a beamwidth $W$ in the $y$-plane.
Then, $\int d\mathbf{k\rightarrow}\frac{2\pi}{W}\sum_{k_{y}}\int
dk_{x}$ and $\hat{b}(  \mathbf{k})  \rightarrow(
\frac{W^{1/2}}{2\pi})  \hat{b}(  k_{x})  $, which leads to 
\bqa
\hat{\mathbf{A}}\left(  \mathbf{r},t\right)  &=&   \frac{1}{2\pi}\int
dk_{x}\sqrt{\frac{\hslash}{2\varepsilon_0 W\omega N}}
{\bm \phi}\left(z,k_{x}\right)  e^{i\left(
k_{x}x\mathbf{-\omega} t\right)  }\hat{b}(
k_{x}) \nonumber \\
&&+~\text{H.c.},
\eqa
with the spatial mode functions ${\bm \phi}(z,k_{x})$ given in Eqs.~(\ref{MF})-(\ref{MF3}).
Finally, converting to the frequency domain using $k_{x}=v_{g}^{-1}\omega$, $dk_{x}=v_{g}%
^{-1}d\omega$ and $\hat{b}(  k_{x})  \rightarrow\sqrt{v_{g}%
}\hat{b}(  \omega)$, where $v_{g}(  \omega)
=\partial\omega/\partial k_{x}$ is the group velocity, we have the vector potential for quantized surface plasma waves, now surface plasmon polaritons (SPPs), given by
\begin{align}
\hat{\mathbf{A}}_{\text{SPP}}^{\text{TE/TM}}({\bf r},t)=  &
\frac{1}{2\pi}\int d\omega\sqrt{\frac{\hslash}{2\varepsilon_0 Wv_{g}\omega
N^{\text{TE/TM}}}}\nonumber\\
&  \times\bm{\phi}^{\text{TE/TM}}(z,\omega)
e^{-i\omega(  t-x/v_{g})  }\hat{b}(  \omega)
+\text{H.c.} \nonumber
\end{align}
\begin{figure}[t]
\centerline{\psfig{figure=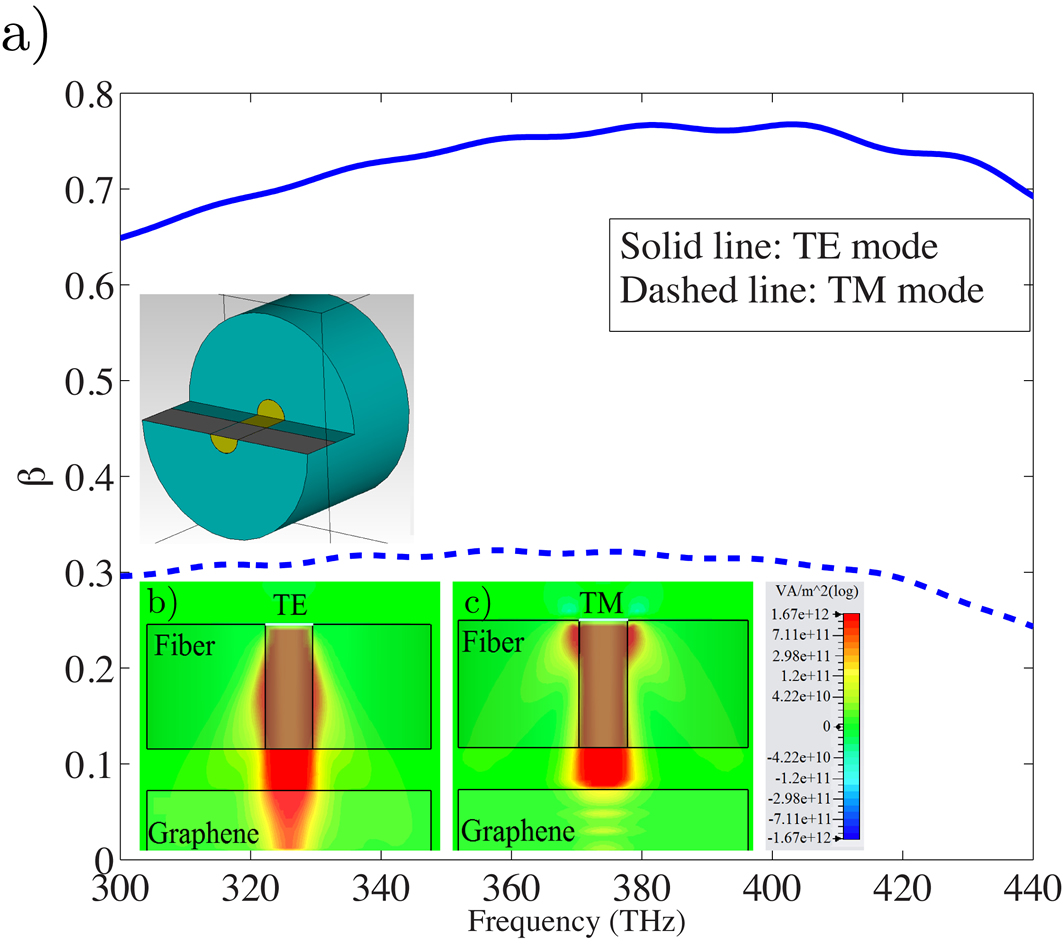,width=7.5cm}}
\caption{End-fire coupling of photons to SPPs on graphene using a fiber. Inset shows the configuration considered. {\bf (a)}: Transmission coefficient $\beta$. {\bf (b)}: Power distribution for TE excitation of graphene strip showing good coupling to a graphene TE SPP. {\bf (c)}: Power distribution for TM excitation of graphene strip showing no excitation of the graphene TM SPP.
The cladding has radius 3 $\mu$m and $\varepsilon_{\rm cladding}=1.16$, the core has radius 500 nm and $\varepsilon_{\rm core}=1.45$, the depth of the removed-region is 3.025 $\mu$m, and the graphene is biased at $\mu_c=0.02$ eV, with $T=0$ K, and $\tau=0.1$ ps ($\text{Im}(\sigma)<0$, allowing only TE mode propagation).
}
\label{fibergraphene}
\end{figure} 

%%%%%%%%%%%%%%%%%%%%%%%%%%%%%%%%%%%%%%%%%%%%%%
%%%%%%%%%%%%%%%%%%%%%%%%%%%%%%%%%%%%%%%%%%%%%%
%%%%%%%%%%%%%%%%%%%%%%%%%%%%%%%%%%%%%%%%%%%%%%

\section{Photon-to-SPP Coupling Model}

With the graphene SPPs quantized we now introduce the coupling model between photons and SPPs in order to investigate the efficiency of the excitation process at the quantum level. As described in Refs. \cite{TLLBPZK2008,BTLLK2009}, within a linear response
regime the coupling of photons to SPPs can be described in the
Heisenberg picture by a unitary transformation matrix
\begin{equation}
\left[
\begin{array}
[c]{c}
\hat{a}_{\text{out}}(  \omega) \\
\hat{b}_{\text{out}}(  \omega)
\end{array}
\right]  =\left[
\begin{array}
[c]{cc}
\gamma(  \omega)  & \beta(  \omega) \\
-\beta^{\ast}(  \omega)  & \gamma^{\ast}(  \omega)
\end{array}
\right]  \left[
\begin{array}
[c]{c}
\hat{a}_{\text{in}}(  \omega) \\
\hat{b}_{\text{in}}(  \omega)
\end{array}
\right],  \label{in_out}
\end{equation}
where $\vert \gamma(  \omega)  \vert ^{2}+\vert
\beta(  \omega)  \vert ^{2}=1$ and $\hat{a}(
\omega)  $ is an annihilation operator for the photon field which,
together with $\hat{a}^{\dag}(  \omega)  $, satisfies the bosonic
commutation relation $[  \hat{a}(  \omega)  ,\hat
{a}^{\dag}(  \omega^{\prime})  ]  =\delta(
\omega-\omega^{\prime})  $. In the following, we will use the coupling
coefficient, $g( \omega)  =e^{i\arg\beta(  \omega)
}\sin^{-1}\vert \beta(  \omega)  \vert$, defined in terms of the transmission coefficient, $\beta (\omega)$, from Eq.~(\ref{in_out}). The coupling coefficient appears in the interaction Hamiltonian for the system, given by $\hat{H}_{int}=i\hbar \int d\omega [g(\omega)\hat{a}^{\dagger} \hat{b} - g^{*}(\omega)\hat{a} \hat{b}^{\dagger}]$. Here, perfect
coupling corresponds to $g(\omega)=\pi/2$, which provides the complete transfer of a given quantum state of a photon to a SPP. 
\begin{figure}[b]
\centerline{\psfig{figure=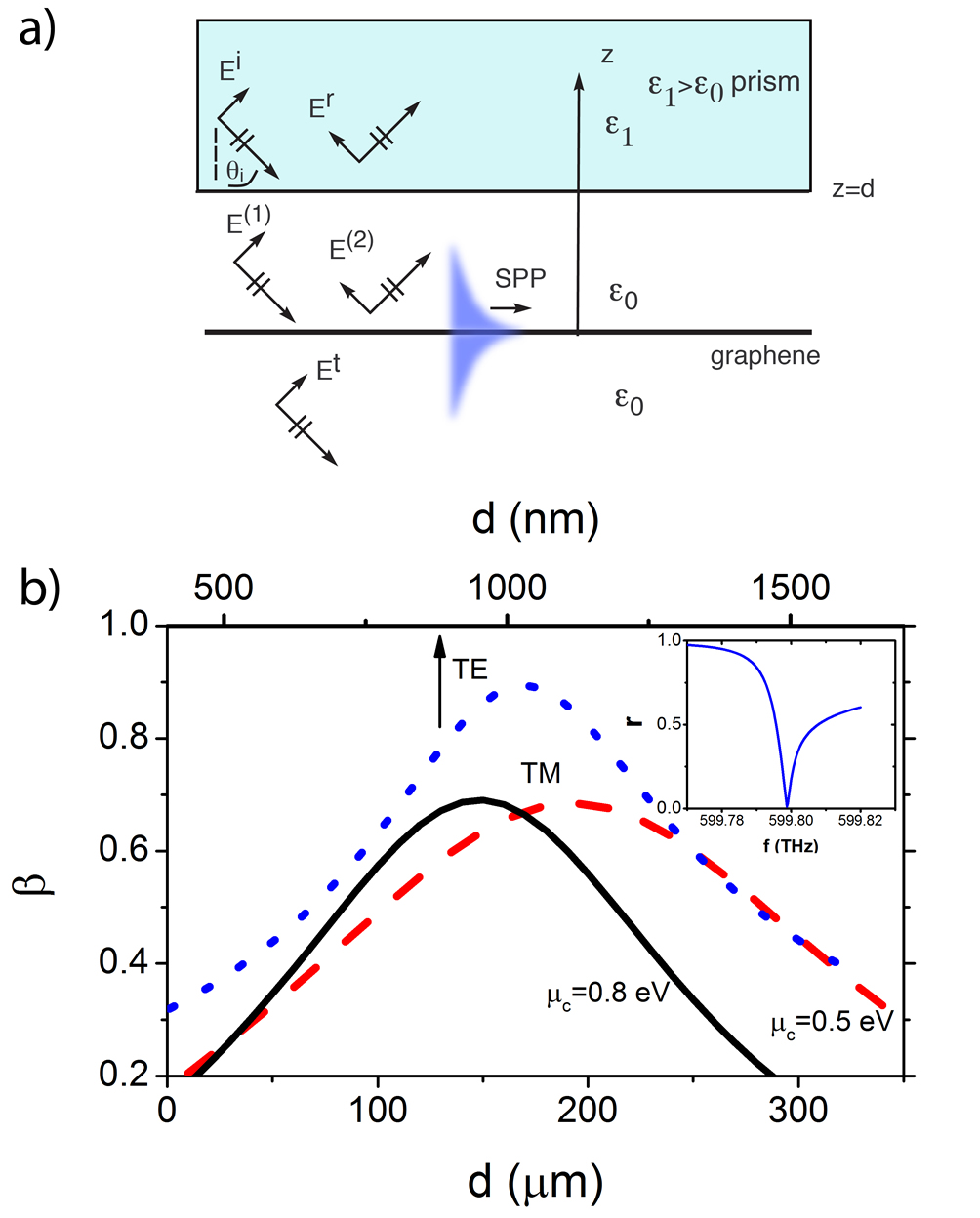,width=7cm}}
\caption{Prism coupling of photons to SPPs on graphene. {\bf (a)}: The coupling system considered. {\bf (b)}: Transmission coefficient for SPPs as a function of prism-graphene spacing $d$. TE mode (dotted, prism-graphene spacing given by top horizontal axis): $\theta_{i}=54.74^{\circ}$, $\varepsilon_{1}=1.5$, $\mu_{c}=\hslash\omega/2=hc/\lambda_0$ ($\mu_{c}=1.24$ eV), $f=600$ THz ($\lambda_0=0.5$ $\mu$m), $T=0$ K, TM mode (dashed, $\mu_c=0.5$ eV, solid, $\mu_c=0.8$ eV, prism-graphene spacing given by bottom horizontal axis): $f=0.81$ THz ($\lambda_0=3.7 \times 10^{-4}$ m), $\theta_{i}=64^{\circ}$, $\varepsilon_{1}=1.5$, $T=300K$. The inset shows the TE reflection coefficient as a function of frequency near the resonance frequency.
}
\label{prismgra}
\end{figure}

In appendix I we provide details of various different coupling scenarios that can provide a range of transformation matrices for the photon-to-SPP transfer. Here we briefly summarize the main results. The first coupling scenario we consider is end-fire coupling, where the photon field from the end of a fiber is evanescently coupled to the field of the graphene SPP~\cite{Bao}. We consider a structure similar to that shown in Fig.~\ref{setup}~(b), where an optical fiber has a center perturbed region with the cladding removed. In order to make the setting more realistic we consider the core of the fiber as being partially removed in order to support the graphene sheet, as shown in the inset of Fig.~\ref{fibergraphene}~(a). Here, photons enter the fiber from the far-right and couple to the graphene SPPs, after which the output is taken on the left hand plane where the structure terminates. We use a numerical FDTD simulation (see Appendix I for details) to obtain the transmission coefficient $\beta$, which is shown in Fig.~\ref{fibergraphene}~(a). The graphene SPPs in this geometry are quantized using the formalism given in the previous section, with appropriate consideration of the asymmetric dielectric media above (air) and below (fiber core support). The FDTD simulation enables the calculation of the overlap between the evanescent photon modefunction and the graphene SPP modefunction. As the modefunctions correspond to the classical wavelike part of the photons and the SPPs, the overlap calculation is essentially a classical calculation with the resulting $\beta$ value entering into the transformation matrix of Eq.~(\ref{in_out}) in order to model the coupling quantum mechanically. Here, the operators are associated with a given spatial modefunction at a specific frequency $\omega$. 

In Fig.~\ref{fibergraphene}~(a) one can see that at high frequencies (where only TE SPPs are supported), good TE SPP coupling can be achieved, whereas TM SPP coupling is significantly reduced. In Fig.~\ref{fibergraphene}~(b) and (c) we show the power distribution for the transfer of the field from the fiber to the graphene sheet for TE and TM modes respectively. One can see that the TE mode can be coupled to well, whereas the TM mode cannot as it is only supported at lower frequencies. In Appendix I we discuss a grating method for efficient end-fire coupling of photons to TM modes at lower frequencies, which is tunable by varying the chemical potential.

The second coupling scenario we consider is prism coupling, where the photon field from below a prism is evanescently coupled to the field of the graphene SPP~\cite{BVP2010,BVP2012,G2012}. The coupling structure is similar to that shown in Fig.~\ref{setup}~(c), whose configuration is shown in more detail in Fig.~\ref{prismgra}~(a). In Appendix I, we provide details of the prism coupling, which, unlike end-fire coupling, we are able to obtain an analytical solution easily and therefore do not need FDTD simulation. In Fig.~\ref{prismgra}~(b), as an example, we show the coupling coefficient $\beta$ as the spacing between the prism and the graphene, $d$, changes for both TE and TM SPPs at a specific frequency for the incoming photon, corresponding to a free-space wavelength of $\lambda_0=500$~nm. In the inset we show the TE reflection coefficient near the resonance of the coupling. It can be seen from Fig.~\ref{prismgra}~(b) that good photon-SPP coupling can be achieved using a prism for both TM and TE SPPs.

In summary, both end-fire and prism coupling methods can provide good photon-to-SPP coupling, with transmission coefficients $\beta>0.7$ ($g>0.77$) for TM and TE modes.

%%%%%%%%%%%%%%%%%%%%%%%%%%%%%%%%%%%%%%%%%%%%%%
%%%%%%%%%%%%%%%%%%%%%%%%%%%%%%%%%%%%%%%%%%%%%%
%%%%%%%%%%%%%%%%%%%%%%%%%%%%%%%%%%%%%%%%%%%%%%

\section{Propagation and Damping Model}

Once the SPP is excited using one of the above methods it propagates along the graphene surface. Here, it is damped by interactions with phonons, impurities, and defects at both the light level (diffraction and radiation at a physical discontinuity) and electron level (intraband electron scattering and interband absorption), as well as with interactions with the thermal bath of field modes. The former (possible diffraction and radiative scattering) is ignored here as we assume an unperturbed graphene surface. On the other hand, both electron-level damping (incorporated in the graphene conductivity
in Eq.~(\ref{GSC})) and thermal interactions cannot be neglected and are accommodated by using a standard
multiple beam splitter model~\cite{L2000,CC1987,JIL1993,TLLBPZK2008,BTLLK2009}, consisting of $N$
quantum beam splitters each with a quantized field mode $\hat{c}%
_{i}(  \omega)  $, $i=1,...,N$. These bath field operators satisfy the bosonic commutation relations $[  \hat
{c}_{i}(  \omega)  ,\hat{c}_{j}^{\dag}(  \omega^{\prime
})  ]  =\delta_{ij}\delta(  \omega-\omega^{\prime})  $. In the continuum limit $N\rightarrow\infty$, $\Delta x\rightarrow0$,
$\hat{c}_{i}(  \omega)  \rightarrow\sqrt{\Delta x}\hat
{c}(  \omega,x^{\prime})  $, and $\delta_{ij}\rightarrow\Delta
x\delta(  x-x^{\prime})  $, and the annihilation operator of the
SPP after travelling a distance $x$ is given by \cite{TLLBPZK2008,BTLLK2009}
\begin{align}
\hat{b}_{\text{out}}(  \omega,x)   &  =e^{ik_x x}\hat{b}_{\text{out}}(  \omega) \label{bout} \\
&  +i\sqrt{2k_0\kappa^{\prime\prime}(  \omega)  }\int_{0}^{x}dx^{\prime}e^{ik_x (
x-x^{\prime}) }\hat{c}(  \omega,x^{\prime})  ,\nonumber
\end{align}
where $k_x = k_x(\omega)=k_0(\kappa^{\prime}(\omega)+i\kappa^{\prime \prime}(\omega))$, and $k_0\kappa^{\prime\prime}(  \omega)  $ is the attenuation factor
for a surface plasma wave (wherein electron-level damping is included). The
continuous field operators obey $[  \hat{c}(  \omega,x)
,\hat{c}^{\dag}(  \omega^{\prime},x^{\prime})  ]
=\delta(  x-x^{\prime})  \delta(  \omega-\omega^{\prime
})  $ and the second term in Eq.~(\ref{bout}) preserves the bosonic nature of
the propagated SPP.

Using the relation $\hat{b}(t)=(2\pi)^{-1/2}\int {\rm d} \omega e^{-i\omega t} \hat{b}(\omega)$, the mean SPP flux at space-time coordinate $(  x,t)$ can be calculated, $f_{out}(x,t)=\langle \hat{b}_{out}^\dag(x,t)\hat{b}_{out}(x,t)\rangle$. For a narrow wavepacket centered at $\omega_{0}$, we have~\cite{TLLBPZK2008}
\begin{equation}
f_{\text{out}}(  x,t)  =e^{-2k_0\kappa^{\prime\prime}x}f_{\text{out}%
}(  t_{R}),
\end{equation}
where $t_{R}=t-x/v_{g}$, with $v_{g}$ being the group velocity at the center
frequency. The mean flux of the quantized SPPs is therefore simply damped by
the classically expected factor $2k_0\kappa^{\prime\prime}$. Using the values of $\kappa^{\prime\prime}$ in Fig.~\ref{conduc} for both TE and TM modes, and the above model for the operator mapping (summarized by Eq.~(\ref{bout})) we are now in a position to further investigate the impact of loss on the transfer of quantum states of SPPs on the graphene surface and quantify the performance of an error-correction code for protecting against this loss.

%%%%%%%%%%%%%%%%%%%%%%%%%%%%%%%%%%%%%%%%%%%%%%
%%%%%%%%%%%%%%%%%%%%%%%%%%%%%%%%%%%%%%%%%%%%%%
%%%%%%%%%%%%%%%%%%%%%%%%%%%%%%%%%%%%%%%%%%%%%%

\section{Robust-to-Loss Quantum State Transfer}

\subsection{Lossy propagation}

In Section III we showed that efficient coupling of incident single photons and graphene SPPs is possible at the quantum level. In this section we now analyze the transfer of more complex photon states to SPP states, and their subsequent propagation. The general setting is the following, the input state is defined as $\vert \Psi \rangle _{\text{in}}=\vert \psi \rangle _{a}\vert 0\rangle _{b}$, where $a$ corresponds to the photon mode and $b$ to the SPP mode (which is initially in the vacuum state). The photon interaction with the SPP via end-fire or prism coupling produces the output state, written as $\ket{\Psi}_{\rm out} = {\cal U}\ket{\Psi}_{\rm in}$, where the unitary transformation ${\cal U}$ is defined by Eq.~(\ref{in_out}). We start by considering the photon input in a superposition of coherent states
\cite{BTLLK2009},
\begin{equation}
\vert \Psi\rangle _{\text{in}}=N(  \vert \alpha
\rangle+\vert -\alpha\rangle)_a  \vert 0\rangle_b,
\label{instate}
\end{equation}
with $\vert \pm\alpha\rangle _{a}=\exp[ -\vert
\alpha\vert ^{2}/2]  \sum_{n=0}^{\infty}(  \pm\alpha)
^{n}/\sqrt{n!}\vert n\rangle $, where $\vert n\rangle $
is a number state and $N=(  2+2e^{-2\vert \alpha\vert^2 }%
)  ^{-1/2}$. Using the transformation matrix from Eq.~(\ref{in_out}) we have
$
\vert \Psi\rangle _{\text{out}}  =N(  \vert \alpha\cos
g\rangle_{a}\vert -\alpha\sin g\rangle_{b}
+ \vert -\alpha\cos g\rangle_{a}
\vert \alpha\sin g\rangle_{b})  
$
where $\beta=\sin g$ and $\gamma=\cos g$. For perfect coupling ($g=\pi/2$) we have
\begin{equation}
\vert \Psi\rangle _{\text{out}}= \vert 0\rangle
_aN( (\vert -\alpha\rangle+\vert \alpha\rangle
)_b
\end{equation}
and the coherent state superposition is transferred perfectly to an SPP superposition. 

In general we are interested in the SPP state itself, so we need to trace out the unobserved photon modes $a$. The density operator for the total photon and SPP system is $\hat{\rho}=\vert
\Psi\rangle _{\text{out\ out}}\langle \Psi\vert $. By tracing out system $a$ we have $\hat{\rho}_{b}={\rm Tr}_{a}\hat{\rho}$, which gives~\cite{BTLLK2009}
\begin{align}
\hat{\rho}_{b} &  = \vert N \vert^{2} (  \vert
\alpha\sin g\rangle \langle \alpha\sin g\vert +\vert
-\alpha\sin g\rangle \langle -\alpha\sin g\vert
\nonumber\\
&  +c_{out}(  \vert \alpha\sin g\rangle \langle
-\alpha\sin g\vert + \vert -\alpha\sin g\rangle \langle
\alpha\sin g\vert )  )  \label{rb}%
\end{align}
where $c_{out}=\exp[-2\vert \alpha\cos g\vert ^{2} ]$. We take this state as the initial mixture (at $x=0$) that we want to propagate a distance $x$ along the graphene in the presence of loss and characterize its decoherence. Using Eq.~(\ref{bout}) and Eq.~(\ref{rb}) one finds~\cite{BTLLK2009,P1990}
\begin{align}
\hat{\rho}_{b} (x)  =  &  \vert N \vert^{2} \left(
\vert -\alpha\sin ge^{-k_0\kappa^{\prime\prime}x}\rangle \langle
-\alpha\sin ge^{-k_0\kappa^{\prime\prime}x}\vert \right. \label{rf}\\
&  ~~~~~~~~ +\vert \alpha\sin ge^{-k_0\kappa^{\prime\prime}x}\rangle
\langle \alpha\sin ge^{-k_0\kappa^{\prime\prime}x}\vert \nonumber\\
&  ~~~~~~~~ +c(x)  (  \vert \alpha\sin ge^{-k_0\kappa
^{\prime\prime}x}\rangle \langle -\alpha\sin ge^{-k_0\kappa
^{\prime\prime}x}\vert \nonumber\\
&  ~~~~~~~~ +   \left.  \vert -\alpha\sin ge^{-k_0\kappa^{\prime\prime}%
x}\rangle \langle \alpha\sin ge^{-k_0\kappa^{\prime\prime}%
x}\vert )  \right) \nonumber
\end{align}
where $c(  x)  =c_{out}\exp [-2\vert \alpha\sin g\vert
^{2} (  1-e^{-2k_0\kappa^{\prime\prime}x}) ]$. Note that at long times (large
$x$) the SPP moves towards the vacuum state, as expected, and at early times
(small $x$), $c(  x)  \simeq c_{out}$, and therefore $\hat{\rho}_{b}(  x)  \simeq\hat{\rho}_{b}(0)$. 

In Section III we showed that good coupling can be achieved using several different coupling methods (end-fire, which has been experimentally demonstrated~\cite{Bao}, and prism coupling~\cite{BVP2010,BVP2012,G2012}), resulting in values of $g\sim0.8$ or higher. In the following investigation of state propagation, rather than link the results to a certain coupling
geometry we assume values of $g$ in a reasonable range.

In order to quantify the effect of loss on the excitation process and subsequent propagation, we use the fidelity $F=\bra{\psi}\hat{\rho}\ket{\psi}$ as a measure of the similarity between two states, one pure $\ket{\psi}$ (ideal) and one mixed $\hat{\rho}$ (damped)~\cite{NC}. When $F=1$ the states are the same and when $F=0$ they are completely orthogonal. In the ideal case, the superposition state $\ket{\psi}$ given in Eq.~(\ref{instate}) will be excited as an SPP and then propagate without loss along the graphene surface. In the realistic case, however, the state $\hat{\rho}$ given in Eq.~(\ref{rf}) will be the state resulting from the non-ideal excitation process and damping. The fidelity thus provides a means to measure how far away in the Hilbert space the damped SPP state is from the ideal (initial photon) state as it propagates along the graphene. 

\begin{figure}[t]
\centerline{\includegraphics[width=8cm]{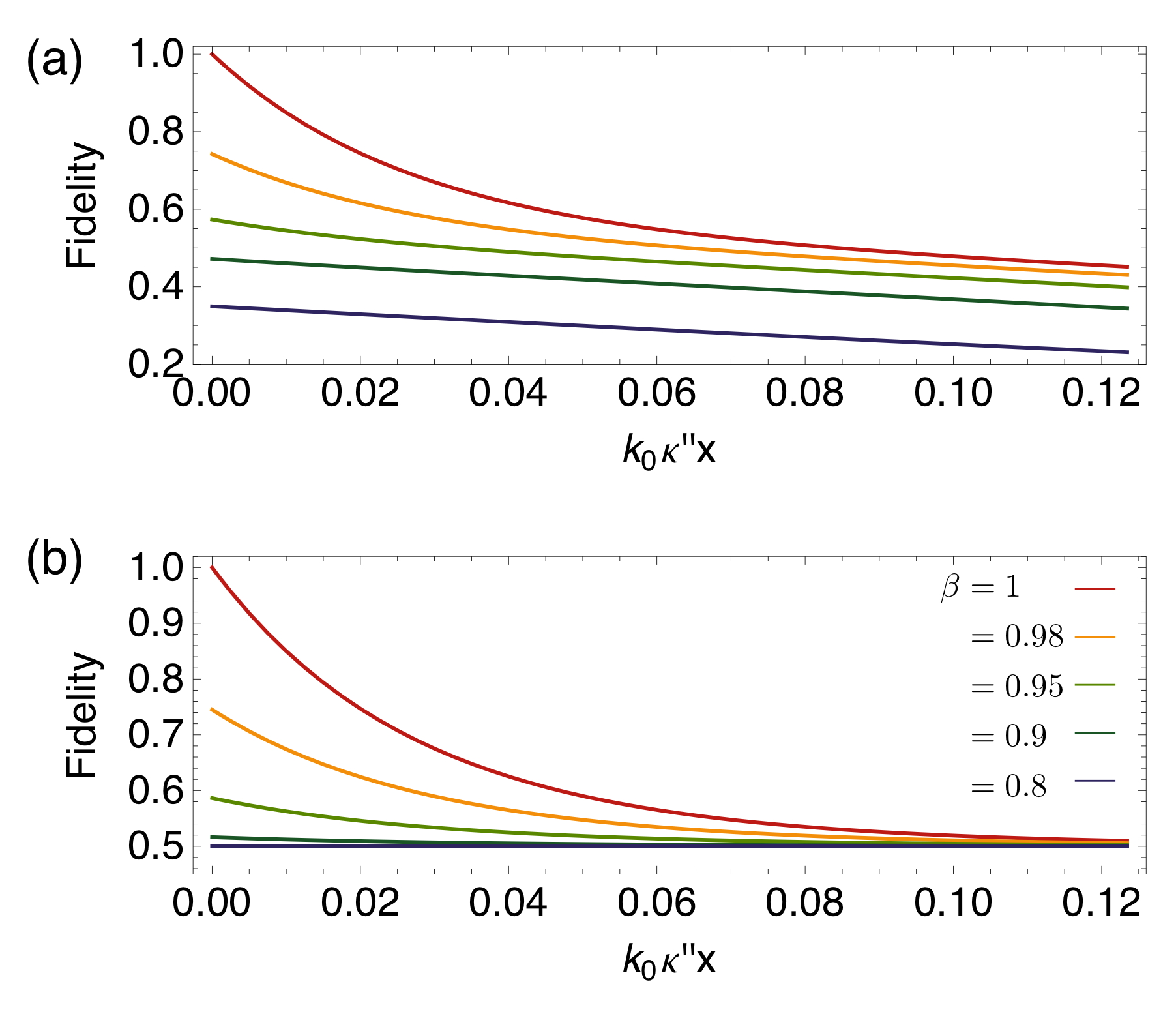}}
\caption{Fidelities of propagating SPPs with respect to the initial state in {\bf (a)} and the initial state with smaller amplitude $\alpha'=\alpha\sin ge^{-k_0\kappa^{\prime\prime}x}$ in {\bf (b)} after excitation with different excitation couplings $\beta=1,0.98,0.95,0.9,0.8$. The initial photon state has $\alpha=3$. The plots in (a) monotonically decrease to zero, while those in (b) approach a finite value when $k_{0}\kappa'' x < {\rm ln}[\sqrt{2}\beta]$, after which they turn and increase to unity as the vacuum contribution becomes significant.}
\label{propagation}
\end{figure} 

In Fig.~\ref{propagation}~(a) we show examples of the fidelities between the initial photon state in Eq.~(\ref{instate}) and the excited SPP states in Eq.~(\ref{rf}) (with different coupling efficiencies) which become damped as they propagate. One can see that the initial fidelities do not start at 1 for the non-ideal coupling cases and the fidelities decay as the SPP propagates, showing the movement of the quantum state further away from the initial ideal state. In Fig.~\ref{propagation}~(b) we show the fidelity of the excited SPP state with respect to the ideal photon state with a smaller amplitude $\alpha'=\alpha\sin ge^{-k_0\kappa^{\prime\prime}x}$. We have included this case as the amplitude of the SPP state is expected to decay as it propagates, thus it is informative to compare it with an ideal photon state that has a decayed amplitude, but importantly has no degradation in its original structure -- it is a pure state with the same fixed positive phase. In this second case the fidelity starts at a slightly higher level and approaches a higher asymptotic value $(=0.5)$ as the damping increases. In Fig.~\ref{propagation}~(a) the asymptotic limit of the fidelity is zero as the SPP state moves toward the vacuum state as a result of dissipation of energy. 
On the other hand, in Fig.~\ref{propagation}~(b) the asymptotic limit is 0.5 as the state we are comparing the SPP state with is matched in terms of its energy, resulting in an effective phase damping of the SPP state from the perspective of the photon state.

From Fig.~\ref{propagation}, it is clear that the excitation process and subsequent damped propagation affect the quality of the quantum state transfer between photons and graphene SPPs. In the next section we will show that by using an error correction strategy, one can protect the superposition state (and more general quantum states) from loss caused by the damping during propagation.

%%%%%%%%%%%%%%%%%%%%%%%%%%%%%%%%%%%%%%%%%%%%%%
%%%%%%%%%%%%%%%%%%%%%%%%%%%%%%%%%%%%%%%%%%%%%%

\subsection{Error-correction code}
In order to provide robust-to-loss propagation of quantum states of SPPs along the graphene waveguide we consider the following code states~\cite{Terhal}
\bqa
\ket{\bar{0}_\pm}&=&\frac{1}{\sqrt{N_\pm}}(\ket{\alpha}\pm \ket{-\alpha}), \\
\ket{\bar{1}_\pm}&=&\frac{1}{\sqrt{N_\pm}}(\ket{i\alpha}\pm \ket{-i\alpha}),
\eqa
where $N_\pm=2(1\pm e^{-2|\alpha|^2})$. For large enough mean photon number $\langle \hat{n} \rangle=|\alpha|^2$ we have that the states $\ket{\pm \alpha}$ and $\ket{\pm i \alpha}$ are orthogonal, and therefore so are the code states. The states $\ket{\bar{0}_+}$ and $\ket{\bar{1}_+}$ form an orthogonal basis, $\vert \langle \bar{1}_{+} \vert \bar{0}_{+}\rangle\vert^{2}\approx0$ when $ \alpha>2$ , representing a code space, in which an arbitrary quantum bit (qubit) can be encoded as
\be
\ket{\Psi}=c_0 \ket{\bar{0}_+}+c_1 \ket{\bar{1}_+},
\ee
where $|c_0|^2+|c_1|^2=1$. This type of encoding has recently been considered theoretically in a cavity scenario using superconducting qubits~\cite{Leg,Mirr} and experimentally demonstrated in Ref.~\cite{Sun}. Here, we extend its application to a waveguide setting. 

When damping occurs during SPP propagation, for a short time interval $\Delta t$ this can be described by the density matrix mapping~\cite{PlenioKnight}
\be
\hat{\rho} \to \hat{\rho}'=\Delta P \hat{\rho}_{jump}+(1-\Delta P) \hat{\rho}_{no-jump},
\ee 
where $\Delta P \ll 1$ is the probability of losing an excitation during time $\Delta t$, $\Delta P=\gamma \Delta t {\rm Tr}(\hat{\rho}\hat{n})=\gamma \Delta t \langle \hat{n}\rangle$, with $\gamma$ the decay rate. Here we have more explicitly $\hat{\rho}_{jump}=\hat{a}\hat{\rho}\hat{a}^\dag/\langle \hat{n}\rangle$ and 
 $\hat{\rho}_{no-jump}= e^{-\frac{\gamma}{2}\Delta t \hat{a}^\dag \hat{a}} \hat{\rho} e^{\frac{\gamma}{2}\Delta t \hat{a}^\dag \hat{a}}/{\rm Tr}[e^{-\frac{\gamma}{2}\Delta t \hat{a}^\dag \hat{a}} \hat{\rho} e^{\frac{\gamma}{2}\Delta t \hat{a}^\dag \hat{a}}]$.
In other words, we can write the evolution of the density matrix during a short time $\Delta t$ as $\hat{\rho} \to\hat{\rho}'=E_0\hat{\rho} E_0^\dag+E_1\hat{\rho} E_1^\dag$, where the Kraus operators for the damping process are $E_0=\sqrt{\gamma \Delta t}\hat{a}$ and $E_1=(1-\frac{\gamma}{2} \Delta t \hat{a}^\dag\hat{a})$, with $e^{-\frac{\gamma}{2}\Delta t \hat{a}^\dag \hat{a}}\simeq (1-\frac{\gamma}{2} \Delta t \hat{a}^\dag\hat{a})$. 
By using the relations
\bqa
\hat{a}\ket{\bar{0}_{\pm}}=\alpha \sqrt{N_{\mp}/N_{\pm}}\ket{\bar{0}_{\mp}}, ~~ \hat{a}\ket{\bar{1}_{\pm}}=i\alpha \sqrt{N_{\mp}/N_{\pm}}\ket{\bar{1}_{\mp}}, && \\
e^{-\frac{\gamma}{2}\Delta t \hat{a}^\dag \hat{a}}\ket{\bar{0}_{\pm}}=e^{\frac{1}{2}|\alpha|^2(e^{-\gamma \Delta t}-1)}\sqrt{N_{\pm,\Delta t}/N_{\pm}}\ket{\bar{0}_{{\pm},\Delta t}}, && \nonumber \\
e^{-\frac{\gamma}{2}\Delta t \hat{a}^\dag \hat{a}}\ket{\bar{1}_{\pm}}=e^{\frac{1}{2}|\alpha|^2(e^{-\gamma \Delta t}-1)}\sqrt{N_{{\pm},\Delta t}/N_{\pm}}\ket{\bar{1}_{{\pm},\Delta t}}, && \nonumber
\eqa
where $N_{\pm,\Delta t}=2(1\pm e^{-2|\alpha e^{-\gamma \Delta t/2}|^2})$ and the decayed states 
\bqa
\hskip-0.5cm \ket{\bar{0}_{\pm,\Delta t}}&=&\frac{1}{\sqrt{N_{\pm,\Delta t}}}(\ket{\alpha e^{-\gamma \Delta t/2}}\pm \ket{-\alpha e^{-\gamma \Delta t/2}}), \\
\hskip-0.5cm \ket{\bar{1}_{\pm,\Delta t}}&=&\frac{1}{\sqrt{N_{\pm,\Delta t}}}(\ket{i\alpha e^{-\gamma \Delta t/2}}\pm \ket{-i\alpha e^{-\gamma \Delta t/2}}),
\eqa
one finds that the code states are mapped as follows
\bqa
\hskip-0.0cm\ketbra{\bar{0}_+}{\bar{0}_+} &\to& \Delta P \ketbra{\bar{0}_-}{\bar{0}_-}+(1-\Delta P)\ketbra{\bar{0}_{+,\Delta t}}{\bar{0}_{+,\Delta t}} \nonumber \\
\hskip-0.0cm\ketbra{\bar{1}_+}{\bar{1}_+} &\to& \Delta P \ketbra{\bar{1}_-}{\bar{1}_-}+(1-\Delta P)\ketbra{\bar{1}_{+,\Delta t}}{\bar{1}_{+,\Delta t}}, \nonumber 
\eqa
and similarly for the off-diagonal terms. The code space and the decayed code space $\{\ket{\bar{0}_{+,\Delta t}} ,\ket{\bar{1}_{+,\Delta t}} \}$ - which remains orthogonal for large enough mean excitation number, can be distinguished from the erred space (spanned by $\ket{\bar{0}_-}$ and $\ket{\bar{1}_-}$, and their decayed versions) by the photon parity operator $\hat{P}=e^{i\pi}\hat{a}^\dag\hat{a}=\sum_n e^{i\pi n}\ketbra{n}{n}=\sum_n(-1)^n\ketbra{n}{n}$. Using the relations $\hat{P}\ket{\alpha}=\ket{-\alpha}$ and $\hat{P}\ket{-\alpha}=\ket{\alpha}$, one finds $\bra{\bar{0}_+}\hat{P}\ket{\bar{0}_+}=\bra{\bar{1}_+}\hat{P}\ket{\bar{1}_+}=+1$ and $\bra{\bar{0}_-}\hat{P}\ket{\bar{0}_-}=\bra{\bar{1}_-}\hat{P}\ket{\bar{1}_-}=-1$. Thus, by measuring the parity continuously within small enough time periods one can determine whether or not an excitation has been lost and correct the state back into the code space. 

Two important points should be mentioned in relation to our application of the above error correction strategy to propagating SPPs in a waveguide setting. First, the correction operations do not need to be performed until the very end of the propagation. This is because the parity checks continuously project the state into either the code space or the erred space, with the state moving between these two subspaces in a cyclical fashion as it propagates, similar to the case described in Refs.~\cite{Leg,Mirr}. At the end of the propagation, after having recorded the outcomes of the sequence of parity checks, we know whether the final state is in the code space or the erred space and we can correct it accordingly to bring it back into the code space, although with a reduced amplitude. This leads to the second important point, which is that once the initial state has propagated a given distance (and undergone many parity check operations), it will have decayed significantly. Therefore the error correction strategy means that we must supply a large enough starting value of $\alpha$ for a desired propagation distance along the graphene surface so that the decayed code states maintain their orthogonality. By doing this we effectively put quantum state transfer and classical state transfer in plasmonics on a level playing field, where one simply increases the intensity in order to transfer information along the plasmonic waveguide.
\begin{figure}[t]
\centerline{\includegraphics[width=6.8cm]{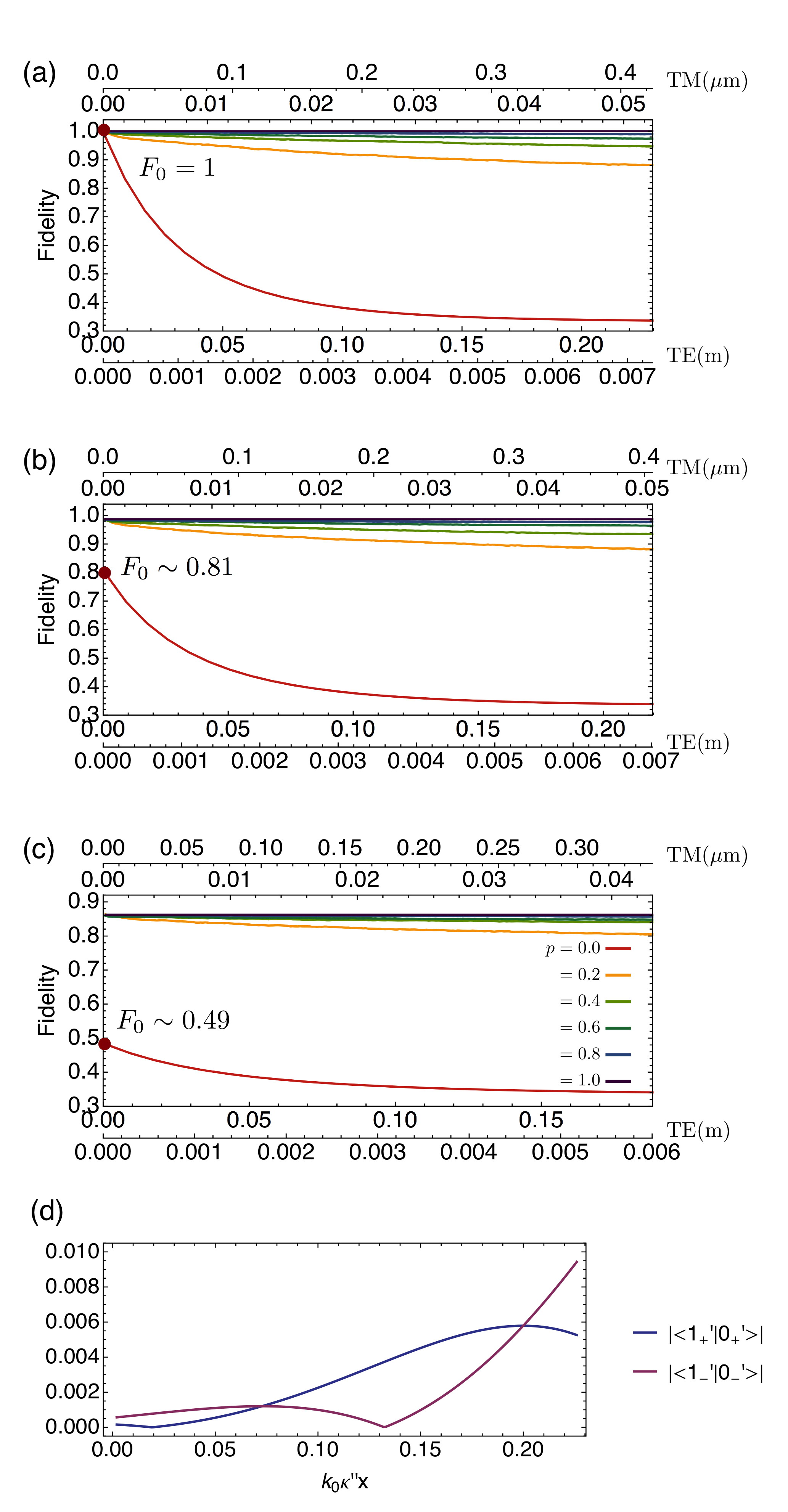}}
\caption{
Average fidelities of error-corrected propagating SPP qubits with increasing parity-check probabilities $p=0,0.2,\cdots,1$ for different excitation couplings: $\beta={\rm sin}[g=\pi/2]$ in {\bf (a)}, $\beta={\rm sin}[g=0.9\times\pi/2]$ in {\bf (b)}, and $\beta={\rm sin}[g=0.8\times\pi/2]$ in {\bf (c)}. $F_{0}$ is the average fidelity just after the excitation process, before the first parity check operation. The fidelities are calculated with respect to the initial state with a smaller amplitude $\alpha'=\alpha\sin ge^{-k_0\kappa^{\prime\prime}x}$ for $\alpha=3$. The propagation length is scaled with respect to TM (TE) modes at T=$300~(0)$ K and $\mu_{c}=1.4~(0.4)$ eV for $\lambda_0=1550$ nm in top (bottom)-upper horizontal axis and $\lambda_0=810$ nm in top (bottom)-lower horizontal axis. The effective wavelengths for TM (TE) modes for $\lambda_0=1550$ nm and $\lambda_0=810$ nm are $\lambda_{\rm eff}(=2\pi/k_{0}\kappa')\sim36.42~(1549.77)$ nm and $\sim 7.23~(810.053)$ nm, respectively. Panel {\bf (d)} checks the orthogonality approximation, $\vert\langle 0_{\pm}' \vert 1_{\pm}' \rangle\vert \leq 10^{-2}$, is well satisfied during a quantum jump simulation for the case of {\bf (c)}, for instance.
}
\label{errorcorrection}
\end{figure} 

In summary, the robust-to-loss encoding for quantum state transfer requires only parity checks to be performed on the SPP as it propagates along the graphene surface. These checks could be realised by incorporating additional circuitry within the graphene sheet, using an ancilla mode (electronic or photonic in origin) and the unitary operation outlined in Ref.~\cite{Sun}, which provides a means of carrying out a non-invasive measurement of the photon parity. The closeness of the parity checks on the surface depends on the rate of loss of the SPP and the speed with which it propagates. For this, we assume the SPP wavepacket containing the quantum state is narrowly centred around the frequency $\omega_0$, such that it propagates with speed $v_G(\omega_0)=v_G$. We then use the relations $\gamma=2k_0\kappa'' v_G$ and $\Delta t=\Delta x/v_G$ to find the corresponding distance $\Delta x$ for the time interval $\Delta t$.

In the following analysis of the performance of the code, we carry out a quantum jump simulation via a Monte Carlo iteration method, tracking $10^4$ individual trajectories of the initial state and summing the final states~\cite{PlenioKnight}. We use the fidelity to quantify the effectiveness of the code as the number of parity checks is modified from the ideal case when it occurs every $\Delta x \ll (2k_0\kappa'' |\alpha|^2)$~m, to the case where no parity checks are made, corresponding to bare graphene propagation. In all cases we consider an initial parity check after the photonic excitation of the SPP in order to put the different cases on an equal footing from the point of excitation. As examples, we show the fidelities (averaged over the single-qubit Bloch sphere) of error-corrected propagation of TM (TE) modes at T=$300~(0)$ K and $\mu_{c}=1.4~(0.8)$ eV for $\lambda_0=1550$ nm and $\lambda_0=810$ nm in Fig.~\ref{errorcorrection} for different excitation couplings in (a), (b), and (c). During propagation, the parity check is performed every $\Delta t/p$ (equivalently $\Delta x/ p$) for a given parity-check probability $p$. 
Note that the proposed scheme corrects the damping effect by flipping over the states in the erred space depending on the outcomes of the parity measurements, leading to a significant increase of fidelity from the initial value $F_{0}$ given by the excitation process.
In Fig.~\ref{errorcorrection}~(d) we show the validity of the orthogonality approximation of the code basis states for the fidelity calculation as the SPP propagates, which we set via $\vert\langle 0_{\pm}' \vert 1_{\pm}' \rangle\vert \leq 10^{-2}$. 

%%%%%%%%%%%%%%%%%%%%%%%%%%%%%%%%%%%%%%%%%%%%%%
%%%%%%%%%%%%%%%%%%%%%%%%%%%%%%%%%%%%%%%%%%%%%%
%%%%%%%%%%%%%%%%%%%%%%%%%%%%%%%%%%%%%%%%%%%%%%

\section{Conclusions}

In this work we investigated the excitation efficiency and impact of loss on SPP propagation in graphene at the quantum level. We considered two different excitation techniques: end-fire and prism coupling. We started by quantizing the transverse-electric and transverse-magnetic SPP modes in graphene, and used this to build a fully quantum model for the excitation process. We then studied various parameter regimes that enabled the excitation of SPPs by photons and found that efficient coupling of single photons to graphene SPPs is possible. We then studied the subsequent propagation of excited quantum states under the effects of loss induced from the electronic degrees of freedom in the graphene. In order to protect the quantum states from loss we used a quantum error-correction code and found that the code provides a robust-to-loss mechanism for propagating quantum states of light in graphene over large distances. The results and analysis in this work contribute to the growing field of quantum plasmonics, and to the use of graphene as a flexible alternative to basic metallic materials for supporting SPPs and their quantum applications.

%%%%%%%%%%%%%%%%%%%%%%%%%%%%%%%%%%%%%%%%%%%%%%
%%%%%%%%%%%%%%%%%%%%%%%%%%%%%%%%%%%%%%%%%%%%%%
%%%%%%%%%%%%%%%%%%%%%%%%%%%%%%%%%%%%%%%%%%%%%%

\section*{Acknowledgements}

We thank C. Noh and S.-W. Lee for discussions and helpful comments. This work is based on research supported by the National Research Foundation and Ministry of Education Singapore (partly through the Tier 3 Grant ``Random numbers from quantum processes''), the South African National Research Foundation and the South African National Institute for Theoretical Physics.

%%%%%%%%%%%%%%%%%%%%%%%%%%%%%%%%%%%%%%%%%%%%%%
%%%%%%%%%%%%%%%%%%%%%%%%%%%%%%%%%%%%%%%%%%%%%%
%%%%%%%%%%%%%%%%%%%%%%%%%%%%%%%%%%%%%%%%%%%%%%

\section*{APPENDIX}
\section*{Appendix I: Free-Space Photon to SPP Coupling Models}

In the main text we considered photon-to-SPP quantum state
transfer for several values of the coupling parameter $g$. In this appendix we
provide several coupling models that can be used to achieve good coupling.

%%%%%%%%%%%%%%%%%%%%%%%%%%%%%%%%%%%%%%%%%%%%%%
%%%%%%%%%%%%%%%%%%%%%%%%%%%%%%%%%%%%%%%%%%%%%%

\subsection{End-Fire Coupling}

In the first configuration, we consider end-fire coupling from a fiber (TE) or wire
(TM) to graphene strips. For the TE case it has been experimentally verified
that good fiber-to-graphene coupling exists~\cite{Bao}. THz experiments for
the TM case have not been performed as yet, although far-infrared excitations
of TM SPPs has been shown using an atomic force microscope tip \cite{Fei,Chen}.
\begin{figure}[t]
\centerline{\psfig{figure=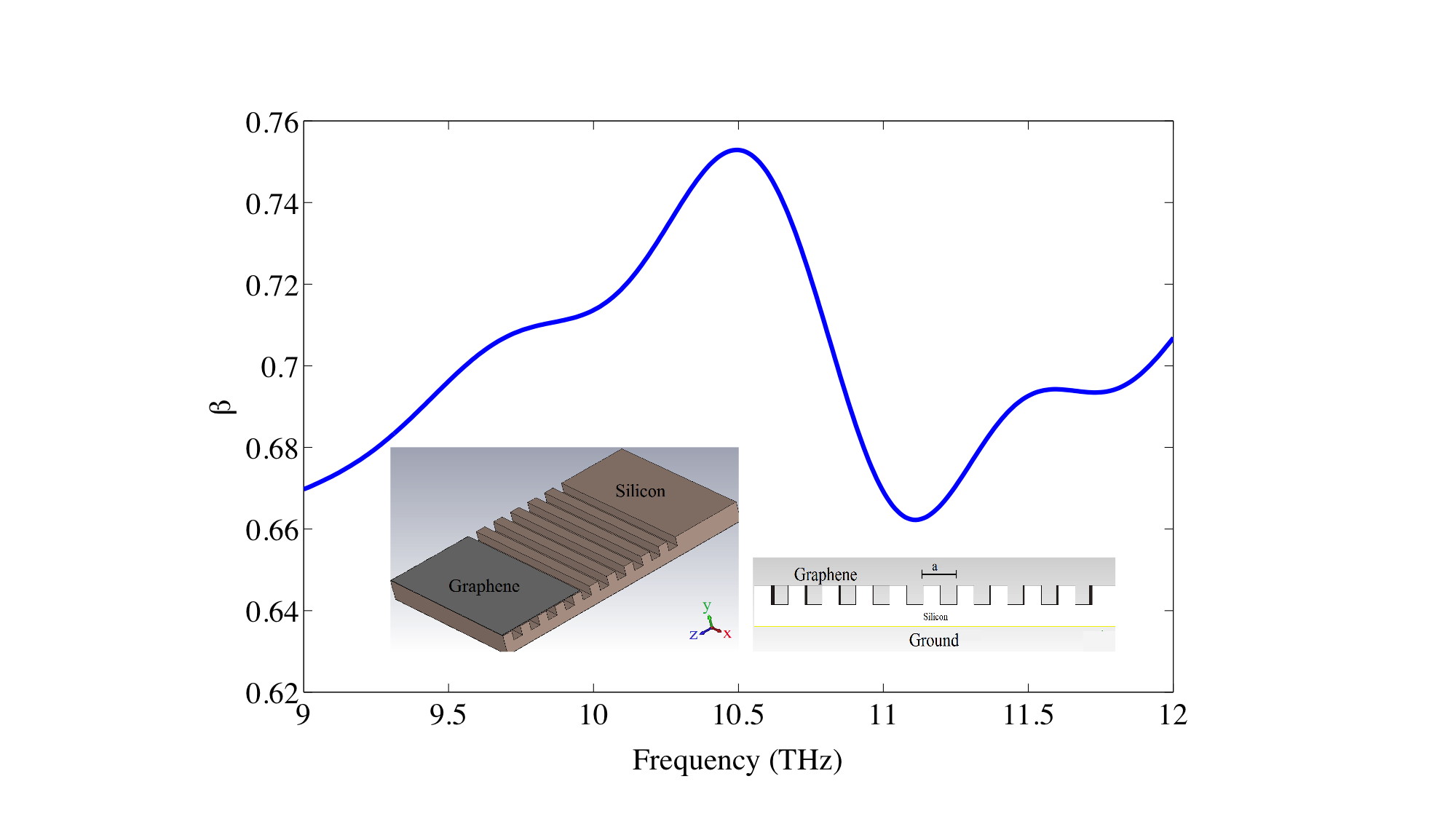,width=9.8cm}}
\caption{Grating end-fire coupling of photons to SPPs on graphene. Inset shows the configuration considered and the main part shows the transmission coefficient $\beta$. 
The grating is silicon, has period 4.717 $\mu$m and depth 2.3 $\mu$m. The width (across) of the graphene region is 39.5 $\mu$m, and $\mu_c=0.152$ eV, $\tau=0.1$ ps at $T=0$ K.
}
\label{gratinggraphene}
\end{figure}

For the TE case we consider a structure similar to the experimental
configuration of Ref.~\cite{Bao}, which consisted of an optical fiber (step-index
core-cladding) with a center perturbed region where the cladding is removed
and the core partially removed, upon which the graphene is located (see inset
to Fig.~\ref{fibergraphene}~(a)). The fiber core is excited at the far-right as the input and the
output is taken at the plane on the left where the structure terminates. In Fig.~\ref{fibergraphene} we show the transmission coefficient $\beta$ obtained using FDTD simulation via CST
Microwave Studio. Very good selectivity for TE polarization is
exhibited, as measured in Ref.~\cite{Bao}. Note, the dimensions of the structure in
Ref.~\cite{Bao} and those used here are somewhat different (see caption for details). Good TE SPP propagation is expected in the frequency range shown, since here $\sigma^{\prime\prime
}<0$ and only a TE SPP propagates. Fig.~\ref{fibergraphene}~(b) and Fig.~\ref{fibergraphene}~(c) show the power distribution on the structure for TE and TM excitation respectively, showing good TE SPP propagation and no TM SPP propagation. 

For the TM case at low THz frequencies, a grating geometry can be used for SPP
excitation. The inset of Fig.~\ref{gratinggraphene} shows the coupling geometry and the main part shows the transmission coefficient $\beta$. At these frequencies $\sigma^{\prime\prime}>0$ and only a TM SPP can propagate.

For both the fiber and grating end-fire coupling methods, the quantization of the graphene SPPs follows the same method as that given in the main text, with appropriate consideration of the asymmetry in the surrounding dielectric media.

%%%%%%%%%%%%%%%%%%%%%%%%%%%%%%%%%%%%%%%%%%%%%%
%%%%%%%%%%%%%%%%%%%%%%%%%%%%%%%%%%%%%%%%%%%%%%

\subsection{Prism coupling}

For prism coupling of photons to SPPs we consider an attenuated total reflection (ATR) set-up
(Otto configuration~\cite{Otto}) to provide a momentum match between the incoming photon
and the graphene SPP. We ignore photon reflection from the top of the prism
(which can be mitigated using impedance matching), and assume that a
plane-wave photon field is incident on the lower air-prism interface at $z=d$,
as depicted in Fig.~\ref{prismgra}~(a). Total internal reflection will result in an evanescent field in the space below the prism, which can couple to the SPP evanescent
field. Since the space below the graphene is vacuum, the field transmitted
into that region is also evanescent. For graphene, prism coupling to SPPs has
been considered classically in Refs.~\cite{BVP2010,BVP2012,G2012}.

To determine the efficiency of SPP excitation we consider the overlap between
the photon field and the SPP field. For the photon
field we assume an incident field having
parallel polarization. In Appendix II we quantize the TE and TM photon modes for the ATR geometry, leading to the quantized vector potential operator for photons having the form
\begin{align}
\hat{\mathbf{A}}_{\text{p}}(  \mathbf{r},t)   &  \propto\int
d\omega(  r(  \omega)  \Psi^{u}(  \mathbf{r}%
,\omega)  \Theta(  z-d)  \nonumber \\
&  + \tau(  \omega)  \Psi^{L}(  \mathbf{r}%
,\omega)  \Theta(  d-z)  )  \hat{a}(
\omega)  N_{\Psi}^{-1/2}(  \omega)  +\text{H.c.}%
,\nonumber
\end{align}
where $N_{\Psi}$ is a normalization parameter having units of length, and
$r(  \omega)$ and $\tau(
\omega)$ are field amplitudes. The factors $\Psi^{u,L}(  \mathbf{r}%
,\omega)  $ are mode functions that depend on the geometry. The
upper-region mode function $\Psi^{u}$ has a standing wave behavior in $z$ and
cannot couple to the graphene plasmons, so that this term can be ignored. The
lower-region mode function $\Psi^{L}(  \mathbf{r},\omega)
=\Psi^{m}(  \mathbf{r},\omega)  (  \Theta(  z)
-\Theta(  z-d)  )  +\Psi^{t}(  \mathbf{r},\omega)
\Theta(  -z)  $ exhibits evanescent behavior and couples energy
into the SPP.

As described in Refs.~\cite{TLLBPZK2008,BTLLK2009}, the transmission coefficient is
the overlap integral between the two fields (photon and SPP), given by
\begin{align}
\beta^{\ast}(  \omega)   &  =\sqrt{1-\vert r (
\omega)  \vert ^{2}}\delta (  \omega-\omega^{\prime})
\delta(  k-k_{x}^{i}) \label{beta}\\
&  \times\frac{1}{\sqrt{N_{\Psi}(  \omega)  }\sqrt{N_{\bm{\phi
}}(  \omega^{\prime})  }}\int dz {\Psi}(  z,\omega
)  \cdot\bm{\phi}(  z,\omega^{\prime})  ^{\ast
}.\nonumber
\end{align}
Here, $\omega$ ($\omega^\prime$) is the radian frequency of the photon (SPP), $k^i_x$ ($k$) is the $x$ component of the photon (SPP), $N_\phi$ is the SPP normalization and the SPP mode functions $\bm{\phi}(  z,\omega^{\prime})$ are given explicitly in Eq.~(\ref{MF}). The form of the ATR mode functions is
given in Appendix II. The transmission coefficient $\beta$ depends on the geometrical
parameters of the prism-graphene system (governing the reflection coefficient
$r$ and the degree of mode overlap), and representative results are presented
in the following discussion.
\begin{figure}[t]
\centerline{\psfig{figure=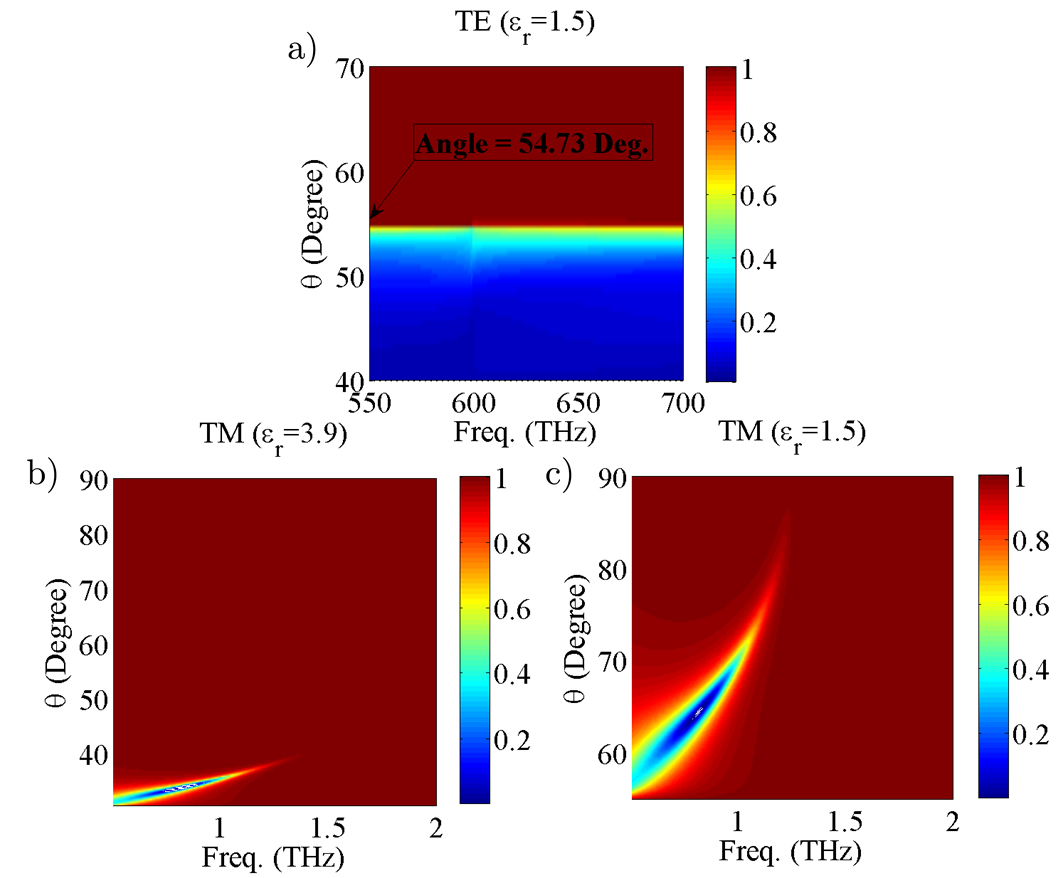,width=8.8cm}}
\caption{Reflectance for the prism-coupling configuration. {\bf (a)}: Reflectance $R=\vert E_{r}\vert ^{2}$ as a function of frequency and incident angle for the TE case, for a prism having $\varepsilon_{1}=1.5$, prism-graphene spacing $d=620$ nm, and $\mu_{c}=1.24$ eV. {\bf (b)}: TM reflectance for $\varepsilon_{1}=1.5$, $d=200$ $\mu$m and $\mu_{c}=0.5$ eV. {\bf (c)}: TM reflectance for $\varepsilon_{1}=3.9$, $d=200$ $\mu$m and $\mu_{c}=0.5$ eV.}
\label{prismref}
\end{figure}

In order to excite SPPs on the graphene surface, the longitudinal wavenumber
of the incident field, $k_{x}^{i}=k_{1}\sin\theta_{i}$, must match the
propagation wavenumber of the SPP given in Eqs.~(\ref{kp})~and~(\ref{kp2}), where $k_1=\sqrt{\varepsilon_1}k_0$ and $\varepsilon_1$ is the relative permittivity of the prism. Therefore, setting
$k_{x}^{\text{SPP}}=k_{1}\sin\theta_{i}$ leads to the matching frequency
$\omega_{0}$ that satisfies
\begin{align}
\sigma(  \omega_{0})  \eta &  =\pm2\sqrt{1-\varepsilon_{1}\sin
^{2}\theta}\text{ \ \ TE}, \nonumber \\
\sigma(  \omega_{0})  \eta &  =\frac{\pm2}{\sqrt{1-\varepsilon
_{1}\sin^{2}\theta}}\text{ \ \ TM}, \nonumber
\end{align}
where $\theta>\theta_{c}$, the critical angle for total internal reflection
(TIR). However, this $\theta$ is only real-valued for lossless graphene
($\sigma=i\sigma^{\prime\prime}$). For the more realistic lossy case one must
find a zero or minimum of the reflection coefficient $r$ at the prism-air
interface.

In Fig.~\ref{prismref}~(a) we show the reflectance $R=\vert r\vert ^{2}$ as a function of
frequency and incidence angle for the TE case, for a prism having
$\varepsilon_{1}=1.5$, prism-graphene spacing $d=620$ nm, and $\mu_{c}=1.24$
eV. The critical angle for TIR is $\theta_{c}=\sin^{-1}(  1/\sqrt
{\varepsilon_{1}})  =54.736^{\circ}$ (for $\theta<\theta_{c}$ the SPP is
not excited). Since for the TE mode, which is loosely confined to the graphene
surface, $k_{x}^{\text{SPP}}\simeq k_{0}$, a match is found for $\varepsilon
_{1}\sin^{2}\theta\simeq1$, which occurs very close to the critical angle. As
a result, the contour of the coupling angle is a horizontal line near
$\theta\simeq\theta_{c}$.

For the TM case at low THz frequencies, Fig.~\ref{prismref}~(b) shows the reflectance for
$\varepsilon_{1}=1.5$, $d=200$ $\mu$m and $\mu_{c}=0.5$ eV, and Fig.~\ref{prismref}~(c)
shows the result for a prism having $\varepsilon_{1}=4$. Compared to the TE
case, considerable dispersion is found, with good matching at low THz
frequencies and angles moderately above the critical angle.

In Fig.~\ref{prismgra}~(b) we show the transmission coefficient $\beta$ as the prism-graphene spacing $d$ changes for TE (upper horizontal axis) and TM (lower horizontal axis) SPPs. The inset shows the TE reflection coefficient versus frequency near resonance, which occurs for $\theta\simeq\theta_{c}$ and $\hslash\omega\simeq2\mu_{c}$. Although the inset only shows the TE case, the reflection coefficient can be reduced to zero for both the TE and TM cases. Despite this, the overlap integral results in $\beta<1$ since the mode functions outside the prism region are not identical to those inside the prism. Since we do not account for reflection from the prism-graphene interface in a rigorous manner, our calculation of beta is valid when the EM field energy inside the prism is negligible.

%%%%%%%%%%%%%%%%%%%%%%%%%%%%%%%%%%%%%%%%%%%%%%
%%%%%%%%%%%%%%%%%%%%%%%%%%%%%%%%%%%%%%%%%%%%%%
%%%%%%%%%%%%%%%%%%%%%%%%%%%%%%%%%%%%%%%%%%%%%%

\section*{Appendix II: ATR Fields and Quantization}

To compute the reflection coefficient in Eq.~(\ref{beta}) for the ATR
geometry, we assume an incident photon field and solve the plane wave
reflection/transmission problem for the prism-graphene geometry shown in Fig.~\ref{prismgra}. We then quantize the resulting fields and obtain the mode functions.

%%%%%%%%%%%%%%%%%%%%%%%%%%%%%%%%%%%%%%%%%%%%%%
%%%%%%%%%%%%%%%%%%%%%%%%%%%%%%%%%%%%%%%%%%%%%%

\subsubsection{TE prism modes}

For the TE case (perpendicular polarization), we assume an incident field in the prism ($z\geq d$) given by
\begin{align*}
\mathbf{E}^{i}(  \mathbf{r})   &  =E_{\perp
}^{i}\hat{\mathbf{y}}e^{i(  k_{x}^{i}x+k_{z}^{i}z)
}e^{-i\omega t}+\text{c.c.},\text{ }\\
\mathbf{H}^{i}(  \mathbf{r})   &  =\frac{k_{z}^{i}\hat{\mathbf{x}}-k_{x}^{i}\hat{\mathbf{z}}}{\omega\mu_{1}}\,E_{\perp}%
^{i}e^{i(  k_{x}^{i}x+k_{z}^{i}z)
}e^{-i\omega t}+\text{c.c.},
\end{align*}
\bigskip where $k_{x}^{i}=k_{1}\sin\theta_{i}$, $k_{z}^{i}=k_{1}\cos\theta
_{i}$ and $k_y^i=0$. The reflected field in the prism is
\begin{align}
\mathbf{E}^{r}(  \mathbf{r})   &  =E_{\perp
}^{r}\hat{\mathbf{y}}e^{i(  k_{x}^{r}x+k_{z}^{r}z)
}e^{-i\omega t}+\text{c.c.}\nonumber \\
\mathbf{H}^{r}(  \mathbf{r})   &  =\frac{
k_{z}^{r}\hat{\mathbf{x}}-k_{x}^{r}\hat{\mathbf{z}}}{\omega\mu_{1}}\,E_{\perp}%
^{r}e^{i(  k_{x}^{r}x+k_{z}^{r}z)
}e^{-i\omega t}+\text{c.c.},\nonumber
\end{align}
with $k_{x}^{r}=k_{x}^{i}$ and $k_{z}^{r}=-k_{z}^{i}$. The field in the middle region (below the prism and
above the graphene, $0\leq z\leq d$) is
\begin{align}
\mathbf{E}^{m}(  \mathbf{r})   &  =(
E_{\perp}^{m1}e^{i(  k_{x}^{m}x+k_{z}^{m}z)  }+E_{\perp}%
^{m2}e^{i(  k_{x}^{m}x-k_{z}^{m}z)  })\hat{\mathbf{y}}e^{-i\omega t} \nonumber \\&\hskip 0.5cm+\text{c.c.}, \nonumber \\
\mathbf{H}^{m}(  \mathbf{r})   &  =\frac{%
k_{z}^{m1}\hat{\mathbf{x}}-k_{x}^{m1}\hat{\mathbf{z}}}{\omega\mu_{1}}\,E_{\perp}%
^{m1}e^{i(  k_{x}^{m}x+k_{z}^{m}z)}e^{-i\omega t}\nonumber\\
&  +\frac{-k_{z}^{m2}\hat{\mathbf{x}}-k_{x}^{m2}\hat{\mathbf{z}}
}{\omega\mu_{1}}\,E_{\perp}^{m2}e^{i(  k_{x}^{m}x-k_{z}^{m}z)}e^{-i\omega t}+\text{c.c.},
\end{align}
where $k_x^m=k_x^i$ and \bigskip\ $k_{z}^{m}=\sqrt{k^{2}-(  k_{x}^{i})  ^{2}}$. The
transmitted field ($z\leq0$) is
\begin{align}
\mathbf{E}^{t}(  \mathbf{r})   &  =E_{\perp
}^{t}\hat{\mathbf{y}}\,e^{i(  k_{x}^{t}x+k_{z}^{t}z)
}e^{-i\omega t}+\text{c.c.},\text{ }\nonumber \\
\mathbf{H}^{t}(  \mathbf{r})   &  =\frac{%
k_{z}^{t}\hat{\mathbf{x}}-k_{x}^{t}\hat{\mathbf{z}}}{\omega\mu_{1}}\,E_{\perp}%
^{t}e^{i(  k_{x}^{t}x+k_{z}^{t}z)
}e^{-i\omega t}+\text{c.c.},\nonumber
\end{align}
with $k_x^t=k_x^i$ and $k_{z}^{t}=k_{z}^{m}$. Enforcing the boundary conditions of Eq.~(\ref{BC}), we
obtain the reflected and transmitted field amplitudes $E_\perp^i$, $E_\perp^r$, $E_\perp^m$ and $E_\perp^t$. The resulting vector potential
has the form
\begin{align}
&  \mathbf{A}(  \mathbf{r},\omega) \nonumber \\
&  =r(  \omega  \Psi^{u}(  \mathbf{r},t)  \Theta(
z-d)  +\tau(  \omega)  \Psi^{L}(  \mathbf{r},t)
\Theta(  d-z)  +\text{c.c.,}\nonumber
\end{align}
where $\Psi^{u,L}(  z,\omega)  $ are mode functions, and $r(
\omega)  =E_{\perp}^{r}/E_{\perp}^{i}$ and $\tau(  \omega)
=e^{ik_{z}^{r}d}E_{\perp}^{t}/E_{\perp}^{i}$ are reflection and transmission
parameters.

\subsubsection{TM prism modes}

For the TM case (parallel polarization), we assume an incident field
\begin{align}
\mathbf{H}^{i}(  \mathbf{r},t)   &  =
\frac{E_{\parallel}^{i}}{\eta_{1}}\hat{\mathbf{y}}e^{i(  k_{x}^{i}x+k_{z}^{i}z)
}e^{-i\omega t}+\text{c.c.},\text{ }\nonumber \\
\mathbf{E}^{i}(  \mathbf{r},t)   &  =\frac{-k_{z}^{i}\hat{\mathbf{x}}+k_{x}^{i}\hat{\mathbf{z}}}{k_{1}}E_{\parallel}^{i}e^{i(
k_{x}^{i}x+k_{z}^{i}z)  }e^{-i\omega t}+\text{c.c.}\nonumber
\end{align}
\bigskip where $k_{x}^{i}=k_{1}\sin\theta_{i}$, $k_{z}^{i}=k_{1}\cos\theta
_{i}$. The reflected field ($z\geq d$) is
\begin{align}
\mathbf{H}^{r}(  \mathbf{r},t)   &  =
\frac{E_{\parallel}^{r}}{\eta_{1}}\hat{\mathbf{y}}e^{i(  k_{x}^{r}x+k_{z}^{r}z)}
e^{-i\omega t}+\text{c.c.},\text{ } \nonumber \\
\mathbf{E}^{r}(  \mathbf{r},t)   &  =\frac{-
k_{z}^{r}\hat{\mathbf{x}}+k_{x}^{r}\hat{\mathbf{z}}}{k_{1}}\,E_{\parallel}^{r}%
e^{i( k_{x}^{r}x+k_{z}^{r}z)  }e^{-i\omega t}+\text{c.c.} \nonumber
,\nonumber
\end{align}
with $k_x^r=k_x^i$ and $k_{z}^{r}=-k_{z}^{i}$. The field in the middle region (below the prism and
above the graphene, $0\leq z\leq d$) is
\begin{align}
\mathbf{H}^{m}(  \mathbf{r},t)   &  =(
\frac{E_{\parallel}^{1}}{\eta_{0}}e^{-ik_{z}^{m}z}+\frac{E_{\parallel}^{m2}%
}{\eta_{0}}e^{ik_{z}^{m}z})\hat{\mathbf{y}}  e^{ik_{x}^{m}x}e^{-i\omega t}%
+\text{c.c.},\nonumber \\
\mathbf{E}^{m}(  \mathbf{r},t)   &  =\frac{-
k_{z}^{m}\hat{\mathbf{x}}+k_{x}^{m}\hat{\mathbf{z}}}{k_{0}}\,E_{\parallel}^{m1}%
e^{i(  k_{x}^{m}x+k_{z}^{m}z)  }e^{-i\omega t}\nonumber\\
&  +\frac{k_{z}^{m}\hat{\mathbf{x}}+k_{x}^{m}\hat{\mathbf{z}}}{k_{0}%
}\,E_{\parallel}^{m2}e^{i(  k_{x}^{m}x-k_{z}^{m}z)  }e^{-i\omega
t}+\text{c.c.,}\nonumber
\end{align}
with $k_x^m=k_x^i$ and $k_{z}^{m}=\sqrt{k_{0}^{2}-(  k_{x}^{i})  ^{2}}$. The
transmitted field ($z\leq0$) is
\begin{align}
\mathbf{H}^{t}(  \mathbf{r},t)   &  =
\frac{E_{\parallel}^{t}}{\eta_{0}}\hat{\mathbf{y}}\,e^{i(  k_{x}^{t}x+k_{z}^{t}z)
}e^{-i\omega t}+\text{c.c.},\text{ }\nonumber \\
\mathbf{E}^{t}(  \mathbf{r},t)   &  =\frac{-
k_{z}^{t}\hat{\mathbf{x}}+k_{x}^{t}\hat{\mathbf{z}}}{k_{0}}\,E_{\parallel}^{t}%
e^{i(  k_{x}^{t}x+k_{z}^{t}z)  }e^{-i\omega t}+\text{c.c.}
,\nonumber
\end{align}
with $k_x^t=k_x^i$ and $k_{z}^{t}=k_{z}^{m}$. Enforcing the boundary conditions of Eq.~(\ref{BC}), we obtain the reflected and transmitted field amplitudes. The resulting vector potential
has the form
\begin{align}
&  \mathbf{A}(  \mathbf{r},\omega) \nonumber \\
&  =(  r(  \omega)  \Psi^{u}(  \mathbf{r},t)
\Theta(  z-d)  +\tau(  \omega)  \Psi^{L}(
\mathbf{r},t)  \Theta(  d-z)  )  +\text{c.c.,}\nonumber
\end{align}
where $\Psi^{u,L}(  \mathbf{r},t)  $ are mode functions, and
$r(  \omega)  =E_{\parallel}^{r}/E_{\parallel}^{i}$ and
$\tau(  \omega)  =\sqrt{k_{z}^{t}/k_{z}^{i}}e^{ik_{z}^{r}%
d}E_{\parallel}^{t}/E_{\parallel}^{i}$ are reflection and transmission
parameters.

%%%%%%%%%%%%%%%%%%%%%%%%%%%%%%%%%%%%%%%%%%%%%%
%%%%%%%%%%%%%%%%%%%%%%%%%%%%%%%%%%%%%%%%%%%%%%

\subsubsection{TE and TM prism mode quantization}

For either the TE or TM polarization the Hamiltonian can be evaluated in the same way as for the graphene quantization described in the main text, leading to%
\begin{equation}
H=\varepsilon_{0}L^{2}\sum_{\bf k}N_{\bf k}\omega^{2}(  A^{t}_{\bf k}A^{t\ast}_{\bf k}+A^{t\ast}_{\bf k}%
A^{t}_{\bf k})  ,
\end{equation}
where $N_{\bf k}$ has units of length, so that the field is quantized using
\begin{align}
A^{t}_{\bf k}   &  \rightarrow\sqrt{\frac{\hslash}{2\varepsilon
_{0}L^{2}\omega N_{\bf k}}}\hat{a}_{\bf k}  ,\\
A^{t\ast}_{\bf k}   &  \rightarrow\sqrt{\frac{\hslash
}{2\varepsilon_{0}L^{2}\omega N_{\bf k}}}\hat{a}^{\dagger}_{\bf k}.
\end{align}
The bosonic annihilation and creation operators $\hat{a}_{\bf k}$ and
$\hat{a}^{\dag}_{\bf k}$ satisfy $[ \hat{a}_{\bf k}
,\hat{a}^{\dagger}_{\bf k^\prime}  ]  =\delta_{{\bf k},{\bf k^\prime}}  $. Assuming a continuum of frequencies, the
quantized potential operator is then%
\begin{align}
\hat{\mathbf{A}}_{\text{p}}(  \mathbf{r},t)  =  &  \frac
{1}{2\pi}\int d\omega\sqrt{\frac{\hslash}{2\varepsilon_{0}L^{2}\omega N}}e^{-i\omega(  t-x/v_{g})  }\hat{a}(  \omega) \\
&  \times(  r(  \omega)  \Psi^{u}(  \mathbf{r}%
,\omega)  \Theta(  z-d)  \nonumber\\
&  + \tau(  \omega)  \Psi^{L}(  \mathbf{r}%
,\omega)  \Theta(  d-z)  )  +\text{H.c.} \nonumber
\end{align}

%%%%%%%%%%%%%%%%%%%%%%%%%%%%%%%%%%%%%%%%%%%%%%
%%%%%%%%%%%%%%%%%%%%%%%%%%%%%%%%%%%%%%%%%%%%%%
%%%%%%%%%%%%%%%%%%%%%%%%%%%%%%%%%%%%%%%%%%%%%%

\end{document}